\documentclass[preprint,12pt,a4paper,onecolumn]{elsarticle}

\usepackage{amssymb}
\usepackage[unicode=true]{hyperref}
\def \CSTE              {\text{CSTE}}
\def \cstebin         {\texttt{cste\_bin()}}
\def \cstebinSCB    {\texttt{cste\_bin\_SCB()}}
\def \cstesurv      {\texttt{cste\_surv()}}
\def \cstesurvSCB   {\texttt{cste\_surv\_SCB()}}
\def \logit             {\text{logit}}
\newcommand{\pkg}[1]{\textbf{\texttt{#1}}}
\usepackage{lmodern}
\usepackage{anyfontsize}
\usepackage{longtable}

\def \CSTE              {\text{CSTE}}
\def \logit             {\text{logit}}
\usepackage{lmodern}
\usepackage{anyfontsize}
\usepackage{graphicx}%
\usepackage{multirow}%
\usepackage{amsmath,amssymb,amsfonts}%
\usepackage{amsthm}%
\usepackage{mathrsfs}%
\usepackage[title]{appendix}%
\usepackage{xcolor}%
\usepackage{textcomp}%
\usepackage{manyfoot}%
\usepackage{booktabs}%
\usepackage{algorithm}%
\usepackage{algorithmicx}%
\usepackage{algpseudocode}%
\usepackage{listings}%
\usepackage{enumitem}

\journal{SoftwareX}

\begin{document}
\renewcommand{\labelenumii}{\arabic{enumi}.\arabic{enumii}}

\begin{frontmatter}

\title{CSTEapp: An interactive R-Shiny application of the covariate-specific treatment effect curve for visualizing individualized treatment rule}


\author[label1,label8,label9]{Yi Zhou\corref{**}}
\author[label1]{Yuhao Deng\corref{**}}
\address[label1]{Beijing International Center for Mathematical Research, Peking University, Beijing, China}
\address[label8]{Division of Mathematics and Informatics, Graduate School of Human Development and Environment, Kobe University, Kobe, Japan}
\address[label9]{Department of Biomedical Statistics, Graduate School of Medicine, The University of Osaka, Osaka, Japan}
\author[label5]{Yu-Shi Tian}
\address[label5]{Graduate School of Pharmaceutical Sciences, Osaka University, Osaka, Japan}

\cortext[**]{These two authors contributed equally to this work.}

\author[label3]{Peng Wu}
\address[label3]{School of Mathematics and Statistics, Beijing Technology and Business University, China}

\author[label4]{Wenjie Hu}
\author[label4]{Haoxiang Wang}
\address[label4]{School of Mathematical Sciences, Peking University, Beijing, China}
\author[label9]{Ewout Steyerberg}
\address[label9]{Scientific Director of the Julius Center, University Medical Center Utrecht, Utrecht, Netherlands}
\author[label1,label6,label7]{Xiao-Hua Zhou\corref{*}}
\address[label6]{Department of Biostatistics, School of Public Health, Peking University, Beijing, China}
\address[label7]{National Engineering Laboratory of Big Data Analysis and Applied Technology, Peking University, Beijing, China}
\cortext[*]{Corresponding author.\ead{azhou@math.pku.edu.cn}}

\begin{abstract}
In precision medicine, deriving the individualized treatment rule (ITR) is crucial for recommending the optimal treatment based on patients’ baseline covariates. The covariate-specific treatment effect (CSTE) curve presents a graphical method to visualize an ITR within a causal inference framework. Recent advancements have enhanced the causal interpretation of the CSTE curves and provided methods for deriving simultaneous confidence bands for various study types. To facilitate the implementation of these methods and make ITR estimation more accessible, we developed CSTEapp, a web-based application built on the R Shiny framework. CSTEapp allows users to upload data and create CSTE curves through simple ``point and click'' operations, making it the first application for estimating the ITRs. CSTEapp simplifies the analytical process by providing interactive graphical user interfaces with dynamic results, enabling users to easily report optimal treatments for individual patients based on their covariates information. {Currently, CSTEapp is applicable to studies with binary and time-to-event outcomes}, and we continually expand its capabilities to accommodate other outcome types as new methods emerge. We demonstrate the utility of CSTEapp using real-world examples and simulation datasets. By making advanced statistical methods more accessible, CSTEapp empowers researchers and practitioners across various fields to advance precision medicine and improve patient outcomes.
\end{abstract}


\end{frontmatter}

\section{Motivation and significance}

\subsection{Covariate-specific treatment effect curve for individualized treatment rule}

In precision medicine, identifying the most effective treatment option for individual patients based on their unique characteristics has become increasingly important. 
This approach, known as an individualized treatment rule (ITR), aims to maximize the benefits of treatments by tailoring them to each patient's unique profile. Traditional methods often focus on average treatment effects across populations, but ITRs allow for more personalized medicine by considering how different patients may respond differently to various treatments.
A review of these methods is presented in Section A of the Supplementary Material.

One of the tools in developing ITRs is the Covariate-Specific Treatment Effect (CSTE) curve, which offers a graphical representation to help visualize how treatment effects vary based on patient characteristics or biomarkers.
The CSTE curve provides a way to understand under what conditions certain treatments are superior for specific subsets of patients. 
Zhou and Ma \cite{Zhou2012} proposed the causal framework for estimating the CSTE curve and established the corresponding simultaneous confidence band (SCB), at a certain confidence level for randomized controlled trials (RCTs) with survival outcomes. 
Building upon this, Ma and Zhou \cite{Ma2017} {considered} scenarios involving multiple treatments. They also proposed methods for constructing SCBs for the CSTE curve using resampling techniques, which helps to understand the uncertainty associated with these estimates.
For RCTs with binary outcomes, Han et al. \cite{KaiShan2017} employed the B-spline method to estimate the CSTE curve and derive the SCB, conditional {on a single biomarker}.
This approach provides a flexible way to model nonlinear relationships between the biomarker and treatment effects.
However, in many clinical trials, there are multiple covariates representing various patient characteristics. 
Handling a large number of covariates can be challenging. To address this, Guo et al. \cite{Guo2021} developed estimation strategies that incorporate variable selection. This method not only estimates the CSTE curve but also identifies which covariates are most influential in determining treatment effects.

Overall, these advancements in estimating CSTE curves and their SCBs provide valuable tools for clinicians and researchers to make informed decisions about treatment options tailored to individual patients' characteristics.
CSTE contributes to the predictive analyses of treatment heterogeneity (PATH) in clinical trials and observational studies \cite{Oleszkowicz2012, Kent2020, Rekkas2023}. CSTE is a risk modeling approach in PATH by considering a linear combination of covariates.

\subsection{Why CSTE curve and CSTEapp are needed}

Compared to other methods for deriving the ITR, the CSTE curve offers a graphical representation of ITR, making it straightforward to select optimal treatments for patients based on their characteristics.
The CSTE curve is more robust against model misspecification {by employing a single-index model with available SCB}, allowing users to assess the variability in the estimated ITR and evaluate the significance of treatment effects conditional on patients’ covariates.
The CSTE curve can be applied to survival outcomes \cite{Zhou2012, Ma2017} and binary outcomes \cite{KaiShan2017, Guo2021}.
However, the estimation procedures for the CSTE curve and the SCB are complicated and sophisticated. 
The lack of software solutions has significantly limited the applicability of the CSTE curve in clinical practice for selecting optimal treatments. 
In order to increase the feasibility and applicability of the CSTE curve to derive the ITR, we developed the interactive web-based application, named CSTEapp, built on the R-Shiny framework and our developed R package \pkg{CSTE}. 
The application includes all existing methodologies that estimate the CSTE curve with the SCB \cite{Zhou2012, Ma2017, KaiShan2017, Guo2021} for studies with binary or survival outcomes.
Since no programming is required, CSTEapp has a lower user burden and is accessible to a wider range of users.

\section{Software description}

CSTEapp is {deployed on a server with an Ubuntu 22.04.4 LTS operating system} and the codes are programmed using R (version 4.3.3).
It is accessible via \url{https://alain003.phs.osaka-u.ac.jp/mephas_web/11cste/} {using any standard internet browser}.
The web-based GUI and statistical computations in CSTEapp are developed using multiple R packages, as listed in Table S1 of the Supplementary Material.
The development of CSTEapp adheres to the Findable Accessible Interoperable Reusable (FAIR) Principles for Research Software \cite{ChueHong2022}, with detailed explanations provided in Tables S2-S3 of the Supplementary Material.

\subsection{Software architecture}

\begin{figure}[!htp] 
\centering
\includegraphics[width=\textwidth]{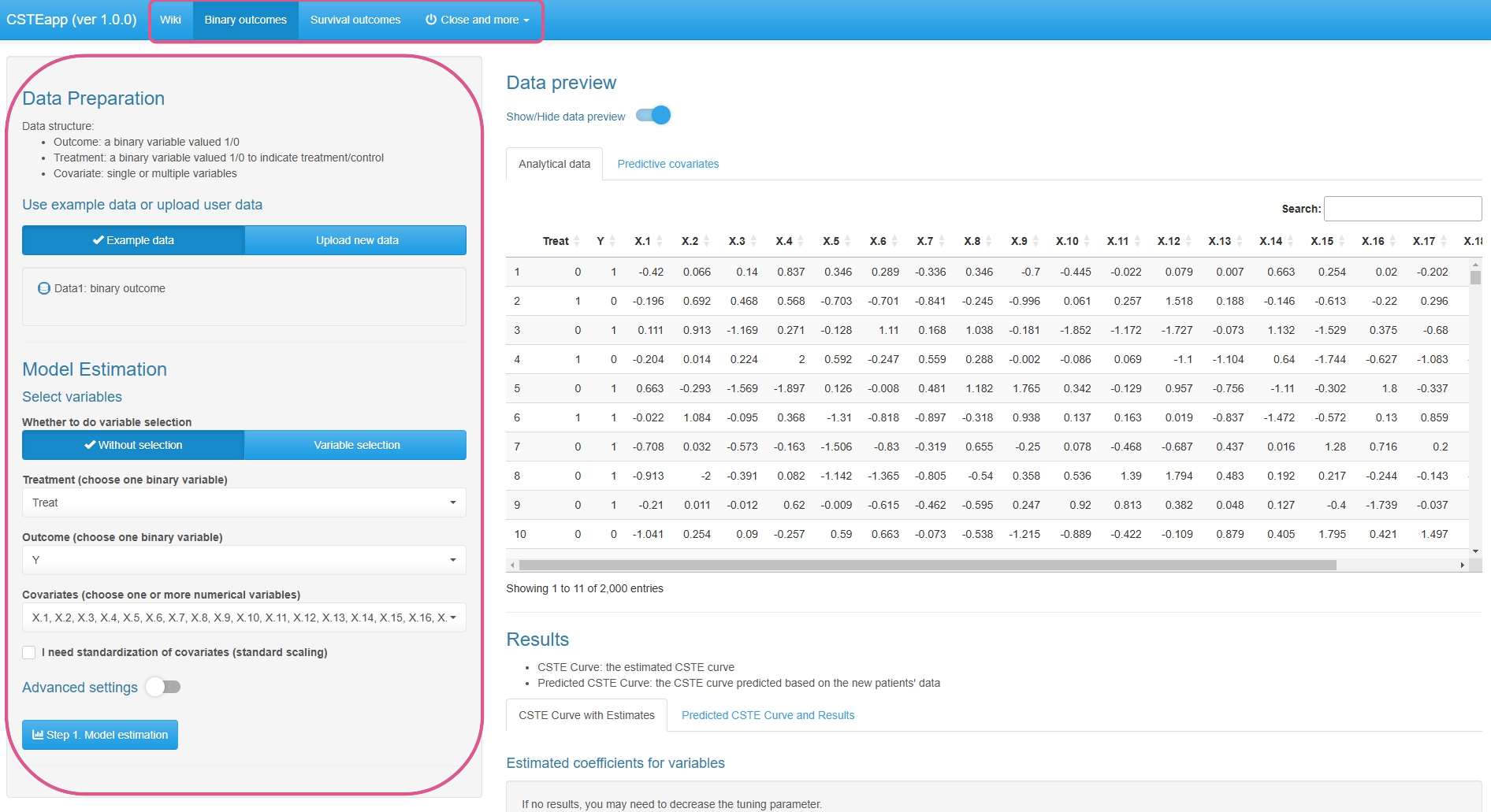}  
\caption{Overview of the layout of CSTEapp. {The red box on the top indicates the navigation bar, and the red box on the left shows the input panel used for analysis.}}
\label{view}
\end{figure}

CSTEapp features three tabs at the top for navigation and one button to close or restart the application (Figure \ref{view}).
The ``Wiki'' tab contains an introduction to CSTEapp and its release history.
The other two tabs navigate to the main parts of the application that estimate the CSTE curves for studies with binary and survival outcomes.
The interfaces of the application are designed in a unified layout divided into a side panel and a main panel.
The left-side panel is used for inputting the analytical dataset and parameters, while the right-side main panel displays the output results (Figure \ref{view}).
The button ``Close and more'' provides a drop-down menu for stopping or restarting the application.

\subsection{Data structure and upload}

CSTEapp includes artificial example datasets to help first-time users familiarize themselves with its operations. If users wish to upload their datasets, they can click the “Upload new data” button and upload the data to CSTEapp  (Figure S1 in the Supplementary Material).
Upon successful upload, the sample data will be replaced, and the new data will be viewed in the “Data preview” section of the main panel.

The CSTE curve applies to intervention studies, such as RCTs with interventions, or observational studies like cohort studies with exposures. It can analyze studies with either binary outcomes or survival outcomes.
For studies with binary outcomes, the data should contain one binary variable of outcome (e.g., $1=$ event; $0=$ non-event), one binary variable of treatment (e.g., $1=$ new treatment; $0=$ control), and at least one covariate variable. 
{Let $Z$ be the exposure, $Y$ be the outcome, and $X$ be covariates. We fit the model $\logit(P(Y=1 \mid X,Z)) = g_1(X^{\top}\beta_1)Z + g_2(X^{\top}\beta_2)$. Then CSTE is given by $g_1(X\beta_1)$.}

For studies with survival outcomes, data should include one continuous variable of time, one binary variable of censoring indicator (e.g., $1=$ event; $0=$ censoring), one covariate variable (i.e., biomarker), and at least one variable of treatment indicators.
{In the case of multi-arm studies, a single treatment indicator can be categorical or multiple binary indicators can be supplied, acting as dummy variables.
Let $\boldsymbol Z$ be the dummy variables of exposures, and $X$ be covariates. We fit the proportional hazards model $\lambda(t\mid X, \boldsymbol Z) = \lambda_0(t) \exp\{\boldsymbol\beta(X)^\top \boldsymbol Z + g(X)\}$. Then the CSTE of the $k$th treatment is given by the $k$th argument of $\boldsymbol\beta(X)$.}

CSTEapp expects a tidy user dataset saved in comma-separated values (CSV) format, where each column of the dataset represents a variable and each row represents an observation (Figures S2-S3 in the Supplementary Material).
If the data are not tidy, the app will not function properly.

\subsection{Implementation of the CSTE curve}

The CSTE curve is derived under the potential outcomes framework \cite{Rubin1974, Splawa-Neyman1990} and represents the difference in the logarithm of odds ratio (lnOR) or the logarithm of hazard ratio (lnHR) between the {exposed and unexposed groups}. 
This allows interpretation of the causal relationship between the exposure and outcomes.

For binary outcomes, the CSTE curve can accommodate large numbers of covariates by using a logit outcome regression model.
This model is fitted by maximizing a penalized log-likelihood function, implemented in C programming.
The sparsity of the coefficient estimates for the covariates, enabling variable selection, can be controlled through the tuning parameter in the penalty function. 
In the estimation process, the B-spline basis functions \cite{Schumaker2007} are used for approximation and implemented by the R function \texttt{bsplineS} in the package \pkg{fda} \cite{fda}.
For survival outcomes, which are commonly right-censored, {the CSTE curve} can handle two or more treatment groups with a single covariate (e.g., biomarker). 
To estimate the CSTE curve, the varying-coefficient proportional hazard regression model {is estimated} by maximizing the logarithm of the local partial likelihood function.
In the case of multiple treatments, the linear combination of the estimated CSTE curves is considered by introducing the contrast vector.
The SCB of the CSTE curve is obtained using the resampling technique \cite{Zhou2012, Ma2017}.
Details of methodologies are summarized in Section E of the Supplementary Material.

\subsection{The CSTE curve creation}

Creating a CSTE curve is designed to be user-friendly, requiring only point-and-click operations after proper data preparation and specification of variables and parameters. The process is detailed in {Section 3} with multiple examples.

To illustrate the use of the CSTE curve, we consider an artificial CSTE curve with binary outcomes, where $Y=1$ indicates death and $Y=0$ otherwise.
Patients are assigned to either a new treatment ($Z = 1$) or a control condition ($Z = 0$). Covariates are represented by $\boldsymbol X$, and $\boldsymbol x$ represents values of these covariates. The CSTE curve, $\CSTE(\boldsymbol x)$, measures the difference in log-odds ratios between treatment groups for {subjects with covariate values $\boldsymbol x$}.

\begin{figure}[!htp] 
\centering
\includegraphics[width=\textwidth]{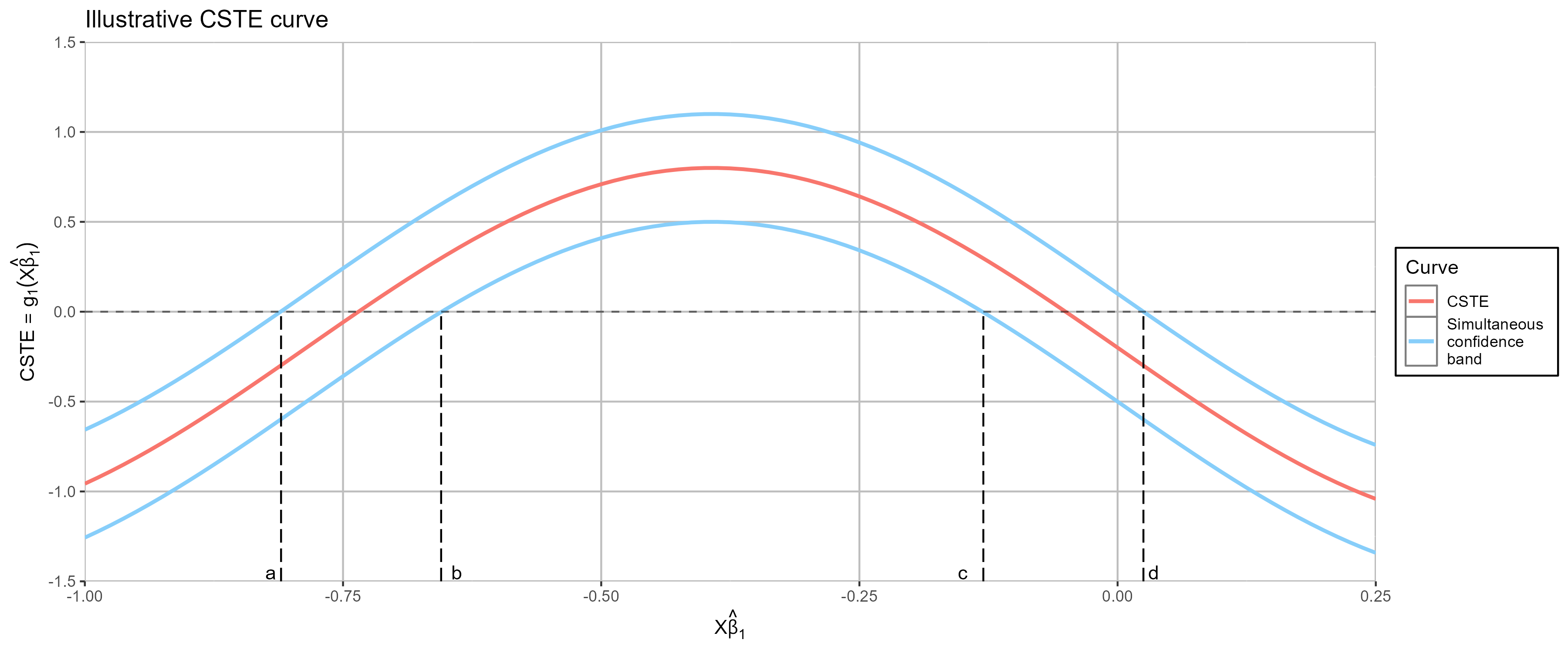}  
\caption{The illustrative CSTE curve along with the simultaneous confidence band (SCB) and cutoff points for binary outcomes. 
Four vertical dashed lines with $\{a,b,c,d\}$ are the locations of cutoff points.}
\label{fig.exam}
\end{figure}

In Figure \ref{fig.exam}, the red curve is the estimated CSTE curve, and the blue curves represent the 95\% SCB.
{The $x$-axis for the CSTE curve is the linear combination of covariates based on estimated coefficients,} denoted by $\boldsymbol x^\top\hat{\boldsymbol\beta}_1$, where $\hat{\boldsymbol\beta}_1$ is the estimated coefficients.
Suppose that $\boldsymbol x^\top\hat{\boldsymbol\beta}_1$ ranges within the interval $(-1,0.25)$. 
The {\it cutoff points} are determined to be the values of $\boldsymbol x^\top\hat{\boldsymbol\beta}_1$ when the SCB intersect with the horizontal line of $\CSTE(\boldsymbol x)=0$ (i.e., points $a$, $b$, $c$, and $d$ in Figure \ref{fig.exam}).
The cutoff points indicate the {\it positive region} of $\CSTE(\boldsymbol x)$ (i.e., $[b,c]$) or {\it negative regions} (i.e., $(-1, a]$ and $[d,0.25)$), helping to recommend the optimal treatment for new patients with their covariates and {the predicted score} $\boldsymbol{x}^\top \hat{\boldsymbol{\beta}}_1$.

The process for selecting the optimal treatment using the CSTE curve is summarized as follows.
\begin{enumerate}[label=Step \arabic*.]
\item Given the observed covariates, estimating the CSTE curve and the corresponding SCB at confidence level $100(1-\alpha)\%$.   
\item Identifying cutoff points where the SCB cross zero and determining positive and negative regions of the CSTE curve.
\item Given a new patient with the covariates $\boldsymbol X=\tilde{\boldsymbol x}$, (1) for binary outcomes, selecting the optimal treatment based on the regions where the predicted $\tilde{\boldsymbol x}^\top\hat{\boldsymbol\beta}_1$ falls; (2) for survival outcomes, selecting the optimal treatment based on the regions where the new value of $\tilde{\boldsymbol x}$ falls.
\end{enumerate}

In the example (Figure \ref{fig.exam}), if a new patient with $\tilde{\boldsymbol x}^\top\hat{\boldsymbol\beta}_1$ falls into the negative regions, $(-1,a]$ or $[d,0.25)$, 
we have 95\% confidence to conclude that the patient will benefit from the new treatment. 
If $\tilde{\boldsymbol x}^\top\hat{\boldsymbol\beta}_1$ falls into $[b,c]$, we have 95\% confidence to conclude that the patient will benefit from the old treatment.
If $\tilde{\boldsymbol x}^\top\hat{\boldsymbol\beta}_1$ falls into $[a,b]$ or $[c,d]$, no significant differences are discovered between these two treatments.

\subsection{Corresponding R package}

For researchers with sufficient experience in R, we developed the R package \pkg{CSTE}. It is available on the Comprehensive R Archive Network (CRAN) and can be installed using the following code:
\begin{verbatim}
R> install.packages("CSTE") 
\end{verbatim} 
The main functions in \pkg{CSTE} are introduced in Table S4 of the Supplementary Material. 

\section{Illustrative examples}

We present four illustrative examples to introduce the usage of CSTEapp in four scenarios. 

\subsection{Example 1: the CSTE curve for binary outcomes without variable selection}

This data comes from The AIDS Clinical Trials Group Study 175 Dataset \cite{Hammer1996}, which aimed to evaluate the efficacy of treatments (zidovudine versus didanosine) in adults infected with human immunodeficiency virus type 1 (HIV-1).
Although the original data included survival outcomes, we used the censoring indicator (\texttt{outcome}) as binary outcomes for illustration purposes. 
In this case, an outcome value of 1 indicates death, and 0 indicates survival.
We focused on determining the optimal treatments between zidovudine and other treatments (\texttt{treat}), where zidovudine was coded as 0 and the other treatments were coded as 1.
Covariates considered included subjects' age (\texttt{age}), weight (\texttt{wtkg}), and measurements of CD4 and CD8 at the baseline or 20 weeks (\texttt{cd40, cd420, cd80, cd820}) as covariates. 
{Since the covariates had varying units, we normalize them before importing the data into the app. CSTEapp also provides a button for normalizing covariates.}
The dataset contained 2,139 subjects. 
For analysis purposes, we randomly sampled 2,100 observations to estimate the model parameters, reserving the remaining 39 subjects for demonstrating prediction capabilities.
In this example, subjects falling into the negative region of the CSTE curve would benefit from the alternative treatment {($Z=1$)}.
Screenshots of the analysis steps are provided in Section G.1 of the Supplementary Material.

To proceed with the analysis, we navigated to the ``Binary outcomes'' tab and clicked the ``Upload new data'' button to upload the dataset.
Once uploaded, the data were displayed in the ``Data preview'' panel on the right side of the interface. 
With the default option set to ``Without selection,'' we proceeded to select the corresponding variables for treatment, outcome, and covariates.
In the advanced settings, we had the option to adjust the number of knots used in the B-spline method. Typically, this can be left at the default value of 2 knots unless there is a specific reason to change it. 
Proceeding with the analysis, we clicked the ``Step 1'' button to estimate the coefficients ($\hat{\boldsymbol\beta_1}$) of the covariates in the CSTE model.
These estimated coefficients were then used in the $x$-axis of the CSTE curve.
Next, we clicked the ``Step 2'' button to estimate the CSTE curve along with its SCB.
This step also activated the prediction panel in the side panel. 
Below this button, advanced options allowed us to adjust the kernel bandwidth in the B-spline method. Increasing the bandwidth would result in a smoother curve. Furthermore, users could modify the significance level for the SCB; commonly, a significance level of 0.05 is used to obtain the 95\% SCB. In this example, we retained the default settings.

\begin{figure}[!htp] 
\centering
\includegraphics[width=\textwidth]{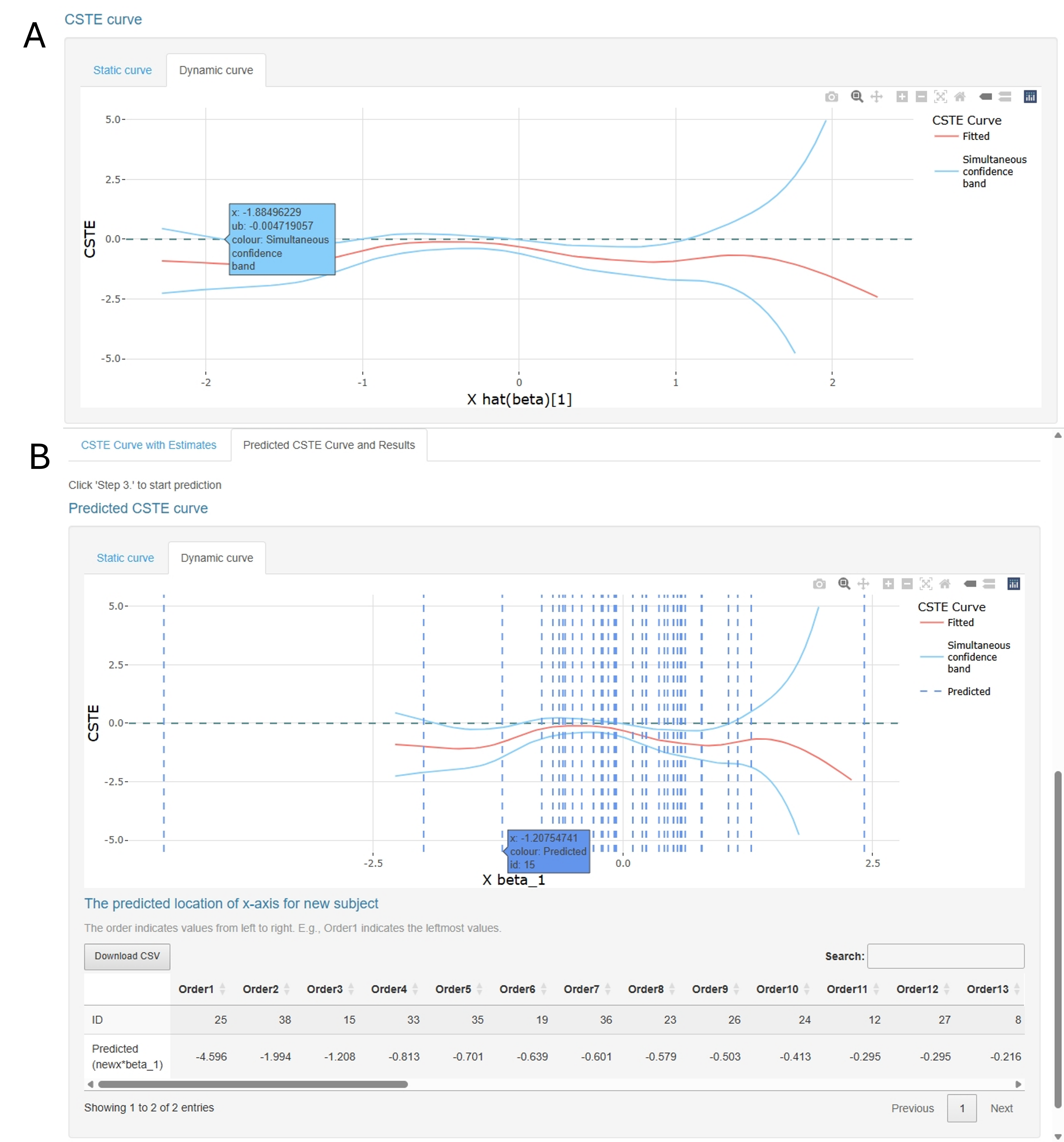}  
\caption{Example 1: estimation (in panel A) and prediction (in panel B) of the CSTE curve.}
\label{p2}
\end{figure}

Finally, by clicking the button to display the curve, we can view the CSTE curve along with its 95\% SCB in both static and dynamic plot formats (Figure \ref{p2}A).
Optional settings allowed us to adjust the limits of the $x$- and $y$-axes.
The static plot could be downloaded by clicking the download button, and the coordinates of the mouse cursor would be displayed when clicking on the plot.
The dynamic plot provided a more interactive way to obtain the precise values from the curves with the mouse cursor hovering over the plot. In this example, we identified the cutoff values and negative regions as $[-1.88, -1.02]$ and $[-0.02, 1.07]$, respectively.

{After completing the estimation}, we uploaded covariates of new subjects in a similar manner and selected the optimal treatment for them. 
By clicking the ``Step 3'' button, we obtained the predicted values for the new subjects.
Then, by clicking the button to show the plot, the predicted values  (i.e., $\tilde{\boldsymbol x}^\top\hat{\boldsymbol\beta}_1$) were displayed as vertical dashed lines over the estimated CSTE curve.
The curves were presented in both static and dynamic formats, with the dynamic curve being more convenient for decision-making (Figure \ref{p2}B), {as it displays the predicted values (locations at $x$-axis shown as x in the floating legend box) and ID (shown as ``id:'' in the floating legend box) of each patient when the mouse cursor hovers over the dashed lines.}
Based on the predicted results {and ID of patients shown in the dynamic plot}, we can be 95\% confident that subjects 15, 9, 14, 4, 21, 29, 1, 7, 17, 16, 18, 6, 5, 32, 22, 20, 28, 34, and 13 would benefit from the alternative treatments. For the remaining subjects, no statistically significant difference was observed between the two treatments.

\subsection{Example 2: the CSTE curve for binary outcomes with variable selection}

We utilized a simulated dataset comprising 20 covariate variables (\texttt{X.1},\texttt{X.2}, \dots, \texttt{X.20}) for illustration purposes.
This dataset consisted of 2,000 observations and contained one binary treatment variable (\texttt{Treat}) and one binary outcome variable (\texttt{Y}).
A subject was assigned to treatment A if \texttt{Treat} had a value of 1 and otherwise a control.
An outcome value of 1 indicated deterioration in health, while 0 indicated no deterioration.
Thus, subjects falling into the negative region of the CSTE curve would benefit more from treatment A.
Screenshots of the analysis steps are provided in Section G.2 of the Supplementary Material.

\begin{figure}[!htp] 
\centering
\includegraphics[width=\textwidth]{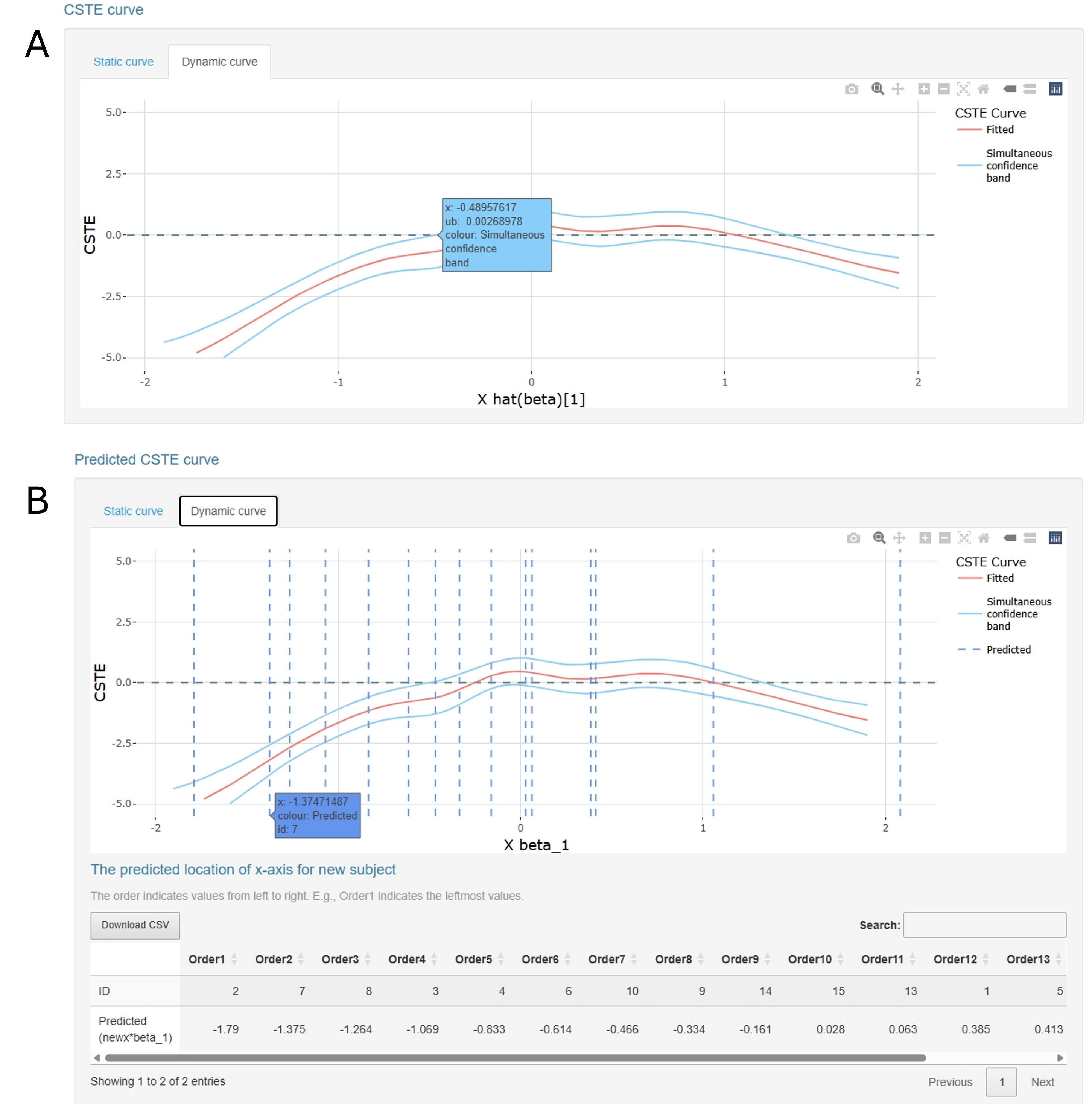}  
\caption{Example 2: estimation (in panel A) and prediction (in panel B) of the CSTE curve.}
\label{eg2-1}
\end{figure}

Since the dataset was already inserted into the application, we used the default ``Example data'' in the data preparation panel.
Next, we clicked the ``Variable selection'' button to select variables and set parameters for the penalized method.
We set the range of tuning parameters from 0.001 to 0.01 with an increment of 0.001; thus, the optimal model would be decided among the models with tuning parameters of 0.001, 0.002, \dots, 0.01.
After clicking the ``Step 1'' button, the optimal estimates of coefficients were estimated at a tuning parameter of 0.008 {with least BIC}.  
We then proceeded to estimate the CSTE curve and generate the static and dynamic plots.
From the dynamic plot, we observed that the negative regions were $(-\infty, -0.49]$ and $[1.32, \infty)$ (Figure \ref{eg2-1}A).

Another simulated dataset with 15 new observations was inserted in the app and used to illustrate the prediction results.
The predicted results were generated by clicking the ``Step 3'' button in the ``Model Prediction'' area and presented in the second tab of ``Results''.
{The predicted values of $\tilde{\boldsymbol x}^\top\hat{\boldsymbol\beta}_1$ were shown as dashed vertical blue lines and ranked by the values. 
The detailed estimation results and the corresponding patient IDs are summarized in the table.}
Based on the dynamic CSTE curve (Figure \ref{eg2-1}B), the 2nd, 7th, 8th, 3rd, 4th, 6th, and 11th subjects could benefit from treatment A.

We validated the estimated CSTE curve with the true CSTE curve using this simulated dataset and found that the estimations were close to the true curves. Further details are presented in Section G.2.1 of the Supplementary Material.

\subsection{Example 3: the CSTE curve for survival outcomes with single treatment}

This example aimed to compare the effects of different transplantation types on leukemia relapse and mortality, using data from leukemia patients who underwent allogeneic stem cell transplantation. 
For illustration purposes, we randomly selected 80 observations, which included a binary variable indicating the type of transplantation  (\texttt{TRANSPLANT}), the overall survival time (\texttt{OST}), a censoring indicator (\texttt{OS}), and a covariate of age (\texttt{AGE}) as the biomarker.
Subjects who underwent human leukocyte antigens (HLA) matched sibling donor transplantation were denoted as \texttt{TRANSPLANT=0} and haploidentical transplantation as \texttt{TRANSPLANT=1}.

\begin{figure}[!htp] 
\centering
\includegraphics[width=\textwidth]{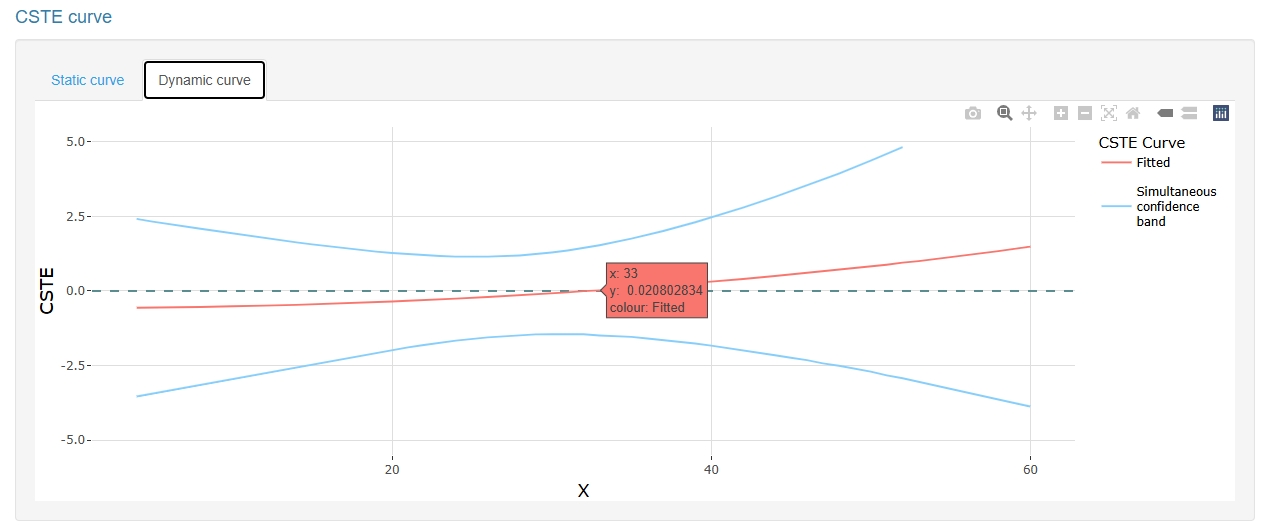}  
\caption{Example 3: estimation of CSTE curve.}
\label{eg3}
\end{figure}

We began by clicking the ``Upload new data'' button and uploading the local CSV file into the application. The uploaded data was shown in the ``Data preview'' panel (Section G.3 of the Supplementary Material).
Since there was only one treatment variable in the data, we clicked ``Single treatment'' and chose the corresponding variables.
With one treatment variable, there was no need to assign a contrast vector.
We then clicked ``Step 1'' to estimate the CSTE curve and generate the plots. 
The advanced settings allowed us to adjust the significance level of the SCB, resampling times for generating the SCB, and the kernel bandwidth. In this case, we used the default settings{, and it takes about 3 seconds to complete the estimations. 
By clicking the button of ``Show/Update the estimated CSTE curve'', the static and dynamic CSTE curves were produced accordingly, showing the relationship between the treatment effect and age.}
By looking at the dynamic CSTE curve (Figure \ref{eg3}), one might conclude that subjects younger than 33 could benefit from haploidentical transplantation (CSTE is negative), while those older than 33 might benefit from HLA-matched sibling donor transplantation. However, the SCB indicated that there was no statistically significant difference between the treatments for subjects of different ages.

\subsection{Example 4: the CSTE curve for survival outcomes with multiple treatments}

In this example, we considered a simulated dataset with multiple treatment options.
{For studies with multiple treatments, we designed two scenarios to accommodate different data formats. The first scenario uses a single treatment variable to represent multiple treatments, as shown in Data 1 provided by CSTEapp. The second scenario uses multiple dummy variables to indicate the different treatments, as illustrated in Data 2 inserted by the app.}
The dataset was inserted as example data, so we skipped the data preparation step.
{Both datasets contained 100 subjects with three treatments: A, B, and C. In Data 1, the treatments were denoted by variable \texttt{Treat}; a subject was assigned to treatment A if \texttt{Treat=0}, to treatment B if \texttt{Treat=1}, and to treatment C if \texttt{Treat=2}.
In Data 2, the treatments were denoted by two dummy variables (\texttt{Treat1} and \texttt{Treat2});
a} subject was assigned to treatment A if \texttt{Treat1=1} and \texttt{Treat2=0}, to treatment B if \texttt{Treat1=0} and \texttt{Treat2=1}, and to treatment C if \texttt{Treat1=0} and \texttt{Treat2=0}. 
The dataset also contained one continuous variable of observed time, one binary variable of censoring indicator, and one covariate of biomarker valued between 0 and 1.

\begin{figure}[!htp] 
\centering
\includegraphics[width=\textwidth]{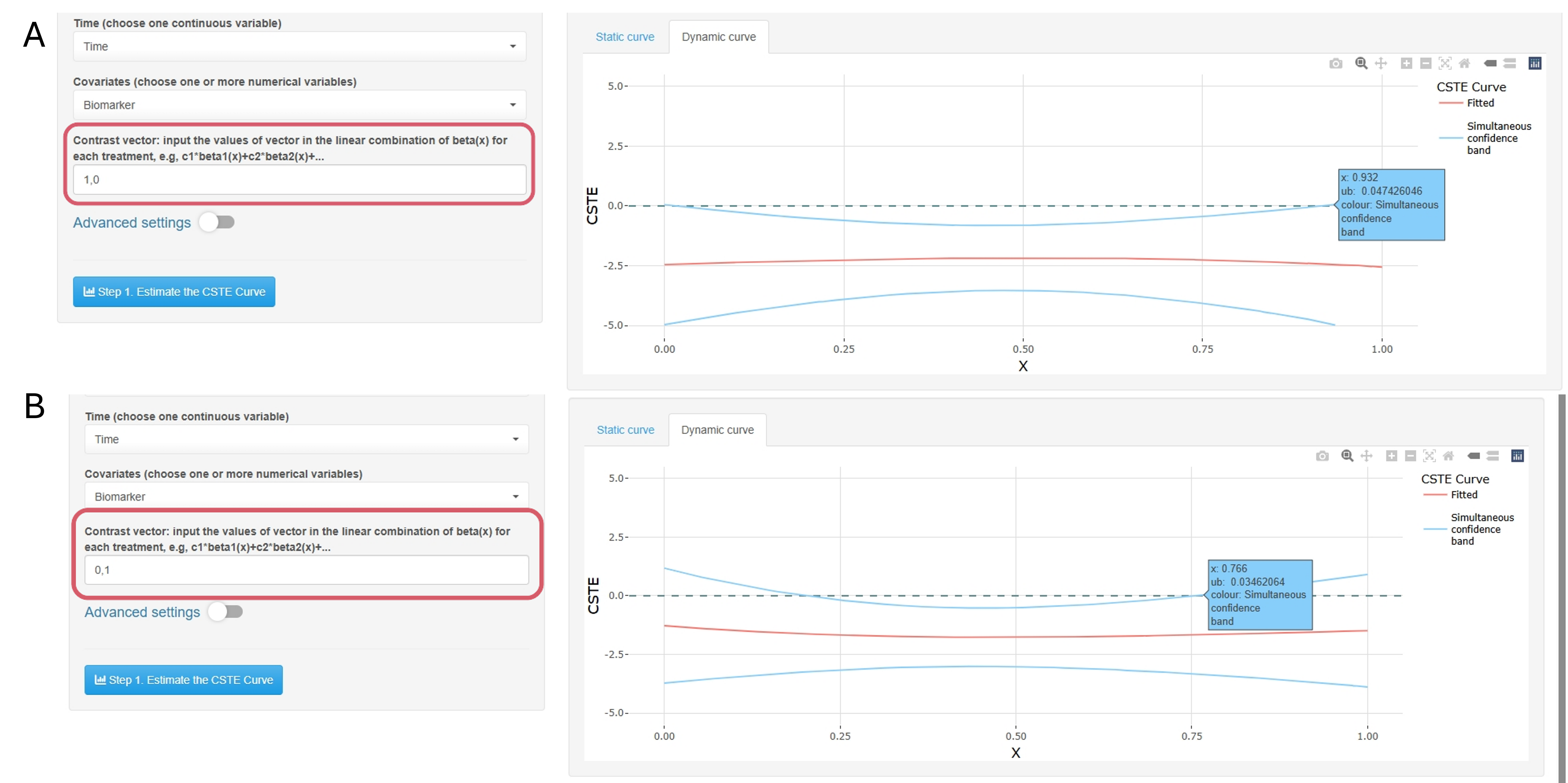}  
\caption{Example 4: estimation of CSTE curve of treatments A versus C (in panel A) and treatment B versus C (in panel B).}
\label{eg4}
\end{figure}

{Using Data 1, we selected the process of ``Multiple treatment coded by a single variable''. Using Data 2 with multiple dummy treatments,} we selected the process of ``Multiple treatment variables'' and chose the corresponding variables (Section G.4 of the Supplementary Material).
To estimate the CSTE curve for different treatments, we should consider the linear combination of the CSTE curve, denoted by $\boldsymbol l^\top\boldsymbol\beta(x)$.
First, we set the contrast vector $\boldsymbol l^\top = (1, 0)$ to estimate the CSTE curve of treatment A versus C.
The dynamic CSTE curve was produced accordingly, with the cutoff point being 0.94.
This implied that we had 95\% confidence that subjects with biomarker values less than 0.94 would benefit from treatment A compared to C (Figure \ref{eg4}A).
Next, we set the contrast vector $\boldsymbol l^\top = (0, 1)$ to estimate the CSTE curve of treatment B versus C (Figure \ref{eg4}B).
After estimating the results, we updated the plots, which showed the cutoff points at 0.20 and 0.80. Thus, we had 95\% confidence to conclude that subjects with biomarker values between 0.20 and 0.80 would benefit from treatment B compared to C.
We validated the performance of the estimated CSTE curve using simulated datasets; the details were presented in Section G.4.1 of the Supplementary Material.

\section{Impact and conclusions}

Estimating the ITR and selecting the optimal treatment regimen have become crucial in precision medicine. 
The CSTE curve visually represents the predictive power of covariates in assessing which patients are likely to benefit more from a treatment compared to others, which contributes to the PATH in clinical trials and observational studies. The CSTE is patient-centered, as it recommends individualized treatment and quantifies the estimation uncertainty through confidence bands. More data from observational studies are used, and the results have a stronger generalizability.
The CSTE curve offers a convenient tool for visualizing the relationship between treatment effects against covariates. However, estimating the CSTE curve requires a strong background in statistics and programming. The practical application of the CSTE curve remains limited in medical decision-making due to the lack of user-friendly software tools.
As the first software designed for deriving the ITR in precision medicine, CSTEapp addresses the issue of software unavailability, thereby facilitating the application of the CSTE curve for optimal treatment selection.
CSTEapp provides a user-friendly interface that allows researchers without extensive coding experience to evaluate the impact of different treatments on various patient subgroups efficiently and reproducibly. With simple point-and-click operations and dynamic and static plots, researchers can evaluate the impact of different treatments on different patients in a reproducible and efficient manner.
We demonstrated the applicability and validated the results of CSTEapp using both real-world examples and simulated datasets. 
We anticipate that CSTEapp will expand the applications of the CSTE curve to diverse fields, including the design of confirmatory trials, policy learning, and personalized product recommendations.

Currently, {CSTEapp is limited to binary and time-to-event outcomes}. In the future, we plan to enhance its capabilities to accommodate other types of outcomes when new methods are developed. 
{For studies with survival outcomes, estimation of the SCB is conducted using bootstrap resampling, which may result in a computational delay of a few seconds. Deriving a theoretical SCB {with an explicit form} is expected to improve the computational efficiency of CSTEapp.
In addition, extending the CSTE curve to survival outcomes with multiple covariates is another potential direction for future updates.}
In summary, CSTEapp offers researchers and practitioners a powerful, user-friendly tool for estimating the CSTE curve, facilitating a broader application in deriving ITRs, and extending its utility across various applications.

\section*{CRediT authorship contribution statement}
{\bf Yi Zhou}: Conceptualization, Formal analysis, Software, Writing – original draft, Writing – review \& editing.
{\bf Yuhao Deng}: Methodology, Software, Writing – original draft, Writing – review \& editing.
{\bf Yu-Shi Tian}: Validation, Writing – original draft, Writing – review \& editing.
{\bf Peng Wu}: Methodology, Software, Writing – review \& editing.
{\bf Wenjie Hu}: Methodology, Software, Writing – review \& editing.
{\bf Haoxiang Wang}: Software, Writing – review \& editing.
{\bf Ewout Steyerberg}: Validation, Writing – review \& editing.
{\bf Xiao-Hua Zhou}: Conceptualization, Supervision, Writing – review \& editing.

\section*{Declaration of competing interest}
The authors declare that they have no known competing financial interests or personal relationships that could have appeared to influence the work reported in this paper.

\section*{Data availability}
The source codes and datasets are released in a GitHub repository \url{https://github.com/mephas/CSTEapp-Rcodes}.
Data in Section 3.1 are accessible via \url{https://archive.ics.uci.edu/dataset/890/aids+clinical+trials+group+study+175}.

\section*{Appendix A. Supplementary data}
The Supplementary Material related to this article can be found online.

\section*{Acknowledgements}
We thank Dr. Yunbei Ma for providing some computer codes on the CSTE curve. 
This work was partially supported by the National Key Research and Development Program of China, Grant No. 2021YFF0901400.

\bibliographystyle{plain}  
\bibliography{refs}  






\appendix
\newpage

\section{Review of methods for estimating the individualized treatment rule}

Several statistical methods have been developed for deriving the ITR based on patients' covariates. 
These methods can be classified in general into two groups and their hybrids.

The first group of methods aims to optimize the average population outcome under the rule of clinically {desired outcomes (e.g., the increase of CD4 counts in HIV patients) or undesired outcomes (e.g., the increase of tumor sizes)}.  
A series of classification algorithm-based methods were proposed by recasting the optimization {as a weighted classification problem} \cite{Zhao2012, Zhang2012, Chen2016, Huang2015, Zhou2017, Zhu2018}, {and some methods have been implemented by R}.
The R function \texttt{wsvm()} (in the \pkg{WeightSVM} package \cite{WeightSVM}) can be used for estimating the optimal treatment in the single decision point setting \cite{Zhang2012};
\texttt{owl()} (in \pkg{DTRlearn2} \cite{DTRlearn2}) and \texttt{rwl()} (in \pkg{DynTxRegime} \cite{DynTxRegime}) implement the outcome weighted learning \cite{Zhao2012} and the residual weighted learning \cite{Zhou2017} algorithms, respectively;    
the methods of Huang \cite{Huang2015}, Chen et al. \cite{Chen2016}, and Zhu et al. \cite{Zhu2018} can be performed by the combination of penalized regression fitting function (\texttt{glmnet()} in \pkg{glmnet} \cite{Friedman2010}), random forest algorithm (\texttt{randomForest()} in \pkg{randomForest} \cite{Liaw2002}), and support vector machines algorithm (\texttt{ksvm()} in \pkg{kernlab} \cite{Karatzoglou2004}).   
However, all these methods have difficulty in establishing valid confidence bands for the ITR \cite{Jiang2020, Laber2019}, making it hard to assess the variation in the estimated ITR.

The second group of methods estimates the heterogeneous treatment effects conditional on patients' covariates \cite{Janes2014, Qian2011, Kang2014, Shi2018}.  
With a single covariate or biomarker, the \pkg{TreatmentSelection} package \cite{TreatmentSelection} was developed for comparing the performance of two treatments \cite{Janes2014}.  
With large numbers of covariate, \texttt{ql()} (in \pkg{DTRlearn2}) implements the Q-Learning algorithm \cite{Qian2011};
\pkg{ada} \cite{ada} implements the boosting iterative algorithm \cite{Kang2014} to reduce the bias caused by the misspecification of the working model;    
\texttt{PAL()} (in \pkg{ITRSelect} \cite{Fan2016,Shi2018}) implements the penalized multi-stage A-Learning algorithm \cite{Shi2018};
\pkg{grf} \cite{grf} estimates the causal forest estimator \cite{Wager2018}.    
These methods either rely on the restrictive assumptions of parametric models or are unable to {handle large numbers of covariates}.

The hybrids of the two groups of methods utilize mean models \cite{Zhang2018, Luckett2020}.
These methods can be implemented by R functions \texttt{rgenoud()} (in \pkg{rgenoud} \cite{Mebane2011,Sekhon1998}) and \texttt{optim()} (in \pkg{stats}) for minimizing the algorithms of C-Learning \cite{Zhang2018} and V-Learning \cite{Luckett2020}, respectively.  
However, these hybrid methods inherit the difficulty of the first group in developing valid confidence bands for the ITR.

\clearpage

\section{R packages used in CSTEapp}
As mentioned in Section 2 of the main text, we present all the R packages used in CSTEapp, as listed alphabetically in Table S\ref{pkgs}.

\begin{table}[!htp]
\caption{The list of R packages used in CSTEapp}
\label{pkgs}
\centering
\fontsize{9}{11}\selectfont
\begin{tabular}{ll}
\hline
Package  name          & Accessible URL                                              \\ \hline
\textit{CSTE}           & https://CRAN.R-project.org/package=CSTE           \\
\textit{DT}             & https://CRAN.R-project.org/package=DT             \\
\textit{fastDummies}             & https://CRAN.R-project.org/package=fastDummies             \\

\textit{ggplot2}        & https://CRAN.R-project.org/package=ggplot2        \\
\textit{htmltools}        & https://CRAN.R-project.org/package=htmltools        \\

\textit{latex2exp}      & https://CRAN.R-project.org/package=latex2exp      \\
\textit{magrittr}      & https://CRAN.R-project.org/package=magrittr      \\
\textit{plotly}         & https://CRAN.R-project.org/package=plotly         \\

\textit{shiny}          & https://CRAN.R-project.org/package=shiny          \\
\textit{shinythemes}    & https://CRAN.R-project.org/package=shinythemes    \\
\textit{shinyWidgets}   & https://CRAN.R-project.org/package=shinyWidgets   \\
\textit{shinycssloaders}   & https://CRAN.R-project.org/package=shinycssloaders   \\
\textit{shinyjs}   & https://CRAN.R-project.org/package=shinyjs   \\

\textit{stats}          & https://CRAN.R-project.org/package=stats          \\
\textit{utils}          & https://CRAN.R-project.org/package=utils      \\ \hline
\end{tabular}
\end{table}

\clearpage

\section{FAIR Principles for Research Software}

As mentioned in Section 2 of the main text, the correspondence of Findable Accessible Interoperable Reusable (FAIR) Principles \cite{ChueHong2022} for Research Software is presented in Table S\ref{fair1}-S\ref{fair2}.

\begin{table}[!htp]
\caption{Correspondence to FAIR Principles for Research Software}\label{fair1}
\fontsize{9}{11}\selectfont
\centering
\begin{tabular}{@{}ll@{}}
\toprule
{\bf F} &  \textit{Software, and its associated metadata, is easy for both humans and machines to find.} \\
Comments & Yes. The URL of CSTEapp (ver 1.0.0) is \\
         & \url{https://alain003.phs.osaka-u.ac.jp/mephas_web/11cste/} \\
         & and metadata \url{https://github.com/mephas/CSTEapp-Rcodes}. \\ \midrule
F1 & \textit{Software is assigned a globally unique and persistent identifier.} \\
Comments & Yes. We assigned version identifiers to CSTEapp, and the current version is 0.9.0. \\ \midrule
F1.1 & \textit{Components of the software representing levels of granularity are assigned distinct}\\
     & \textit{identifiers.} \\
Comments & Yes. We created metadata for CSTEapp (\url{https://github.com/mephas/CSTEapp-Rcodes}). \\ \midrule
F1.2 & \textit{Different versions of the software are assigned distinct identifiers.} \\
Comments & Not applicable. We will assign the different version numbers and record the \\
         & release history once CSTEapp is updated. \\ \midrule
F2 & \textit{This software is described with rich metadata.} \\
Comments & Yes. Metadata can be found via \url{https://github.com/mephas/CSTEapp-Rcodes}. \\ \midrule
F3 & \textit{Metadata clearly and explicitly include the identifier of the software they describe.} \\
Comments & Yes. Metadata is used for CSTEapp (1.0.0). \\ \midrule
F4 & \textit{Metadata are FAIR, searchable and indexable.} \\
Comments & Yes. Metadata can be found via \url{https://github.com/mephas/CSTEapp-Rcodes}. \\ \midrule
{\bf A} & \textit{Software, and its metadata, is retrievable via standardized protocols.} \\
Comments & Yes. CSTEapp can be accessed. \\ \midrule
A1 & \textit{Software is retrievable by its identifier using a standardized communications protocol.} \\
Comments & Yes. CSTEapp can be accessed by HTTPS \\
& (\url{https://alain003.phs.osaka-u.ac.jp/mephas_web/11cste/}). \\ \midrule
A1.1 & \textit{The protocol is open, free, and universally implementable.}\\
Comments & Yes. CSTEapp is free to access. \\ \midrule
A1.2 & \textit{The protocol allows for an authentication and authorization procedure, where necessary} \\
Comments & Currently, there are no specific conditions of access. \\ \midrule
A2 & \textit{Metadata are accessible, even when the software is no longer available.} \\
Comments & Yes. Metadata are stored in \url{https://github.com/mephas/CSTEapp-Rcodes}; \\
         & source code is stored in GitHub. \\ \midrule
\end{tabular}
\end{table}

\begin{table}[!htp]
\caption{Correspondence to FAIR Principles for Research Software (continued)}\label{fair2}
\fontsize{9}{11}\selectfont
\centering
\begin{tabular}{@{}ll@{}}
\toprule
{\bf I} & \textit{Software interoperates with other software by exchanging data and/or} \\
  & \textit{metadata, and/or through interaction via application programming}\\
  &  \textit{interfaces (APIs), described through standards.}\\
Comments & Not applicable. CSTEapp does not interoperate with other software. \\ \midrule
I1 & \textit{Software reads, writes, and exchanges data in a way that meets domain-} \\
   & \textit{relevant community standards.}\\
Comments & Not applicable. CSTEapp needs no exchange of data. \\ \midrule
I2 & \textit{Software includes qualified references to other objects.} \\
Comments & Yes. CSTEapp presented the references about the R packages and data used. \\ \midrule
{\bf R} & \textit{Software is both usable (can be executed) and reusable (can be understood).}  \\
  & \textit{modified, built upon, or incorporated into other software).} \\
Comments & Yes. CSTEapp can be incorporated into other R Shiny-based software. \\ \midrule
R1. & \textit{Software is described with a plurality of accurate and relevant attributes.} \\ 
Comments & Yes. CSTEapp is described in the manuscript; metadata are described in\\
        & the Readme file. \\ \midrule
R1.1 & \textit{Software is given a clear and accessible license.} \\
Comments & Yes. GNU General Public License v3.0 is used for CSTEapp. \\ \midrule
R1.2 & \textit{Software is associated with detailed provenance.} \\
Comments & Yes. The provenance is described in \url{https://github.com/mephas/CSTEapp-Rcodes}. \\ \midrule
R2 & \textit{Software includes qualified references to other software.} \\
Comments & Yes. References to the other R packages are listed. \\ \midrule
R3 & \textit{Software meets domain-relevant community standards.} \\
Comments & Not applicable. There are no specific domain-relevant community standards. \\
         & for Rshiny. \\ \bottomrule
\end{tabular}
\end{table}

\clearpage

\section{Figures of data structure and upload}

\begin{figure}[!htp] 
\centering
\includegraphics[width=0.5\textwidth]{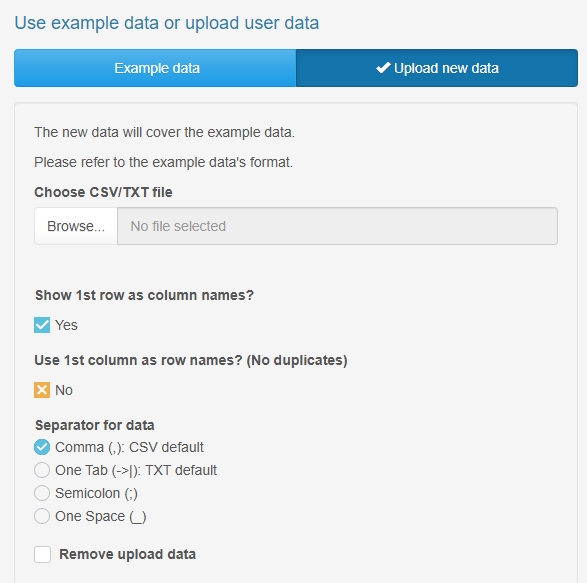}  
\caption{Panel of upload local dataset.}
\label{upload}
\end{figure}

\begin{figure}[!htp] 
\centering
\includegraphics[width=\textwidth]{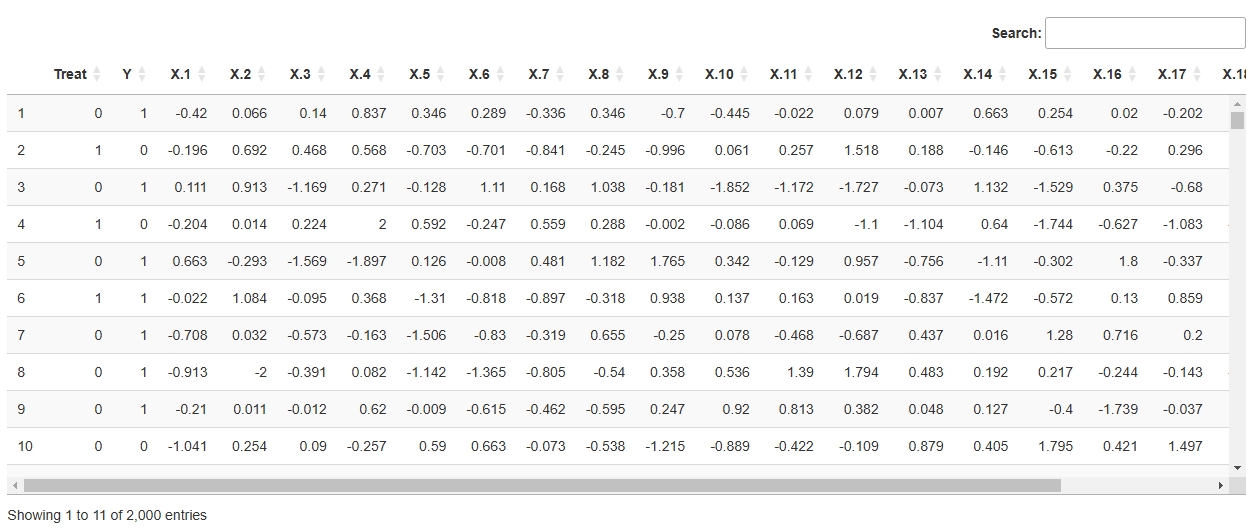}  
\caption{The referential format for data with binary outcomes.}
\label{bdata}
\end{figure}

\begin{figure}[!htp] 
\centering
\includegraphics[width=\textwidth]{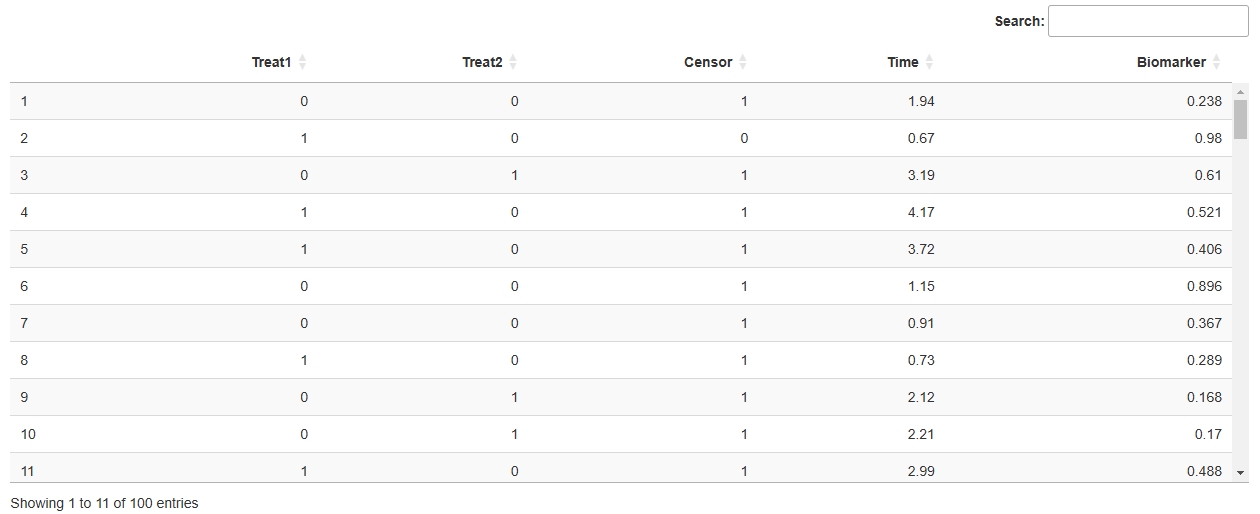}  
\caption{The referential format for data with survival outcomes.}
\label{sdata}
\end{figure}

As mentioned in Section 2.2 of the main text, we presented the screenshot of panels of upload data (Figure S\ref{upload}) and the format of datasets of binary outcomes (Figure S\ref{bdata}) and survival outcomes (Figure S\ref{sdata}).

\clearpage

\section{Methodologies of the CSTE curve}

As mentioned in Section 2.3 of the main text, we present a review of the methodologies for the CSTE curve.

\subsection{CSTE curve for studies with binary outcomes}  

Suppose that, in a study with $n$ subjects, each subject $i$($i = 1, \dots, n$) is assigned to treatment $Z_i$, where $Z_i=1$ if the subject is in the new treatment group and $Z_i=0$ otherwise; 
Let $\boldsymbol X_i=(X_{i1}, \dots, X_{ip})^\top$ be the $p$-dimensional covariates of subject $i$. 
The outcome $Y_i$ is observed to be binary valued.
The data of each subject are assumed to be {independent and identically distributed}; thus, we omit the subscript $i$ hereinafter. 
The treatment effect is estimated under the potential outcomes framework \cite{Rubin1974, Splawa-Neyman1990}, 
Let $Y^1$ and $Y^0$ be the potential outcomes if $Z=1$ and $Z=0$ are received, respectively. 
Under the stable unit treatment value assumption \cite{Basu1980}, the observed outcome $Y$ can be represented as $Y = ZY^1 + (1-Z)Y^0$.  
The common unconfoundedness assumption \cite{RUBIN1976} (i.e., $\{Y^0, Y^1\} \perp Z \mid X$, where $ \perp$ indicates independence) is also imposed for the identification of treatment difference.

With the potential outcomes, the CSTE curve for the binary outcome is defined by 
\begin{align*}
\CSTE(\boldsymbol x)  
= \logit\{E(Y^1 \mid  \boldsymbol X = \boldsymbol x)\}
- \logit\{E(Y^0 \mid  \boldsymbol X = \boldsymbol x)\},
\end{align*}
where $\logit(u) = \log(u) - \log(1 - u)$. 
Under the unconfoundedness assumption, $\CSTE(\boldsymbol x)$ can be re-formulated using the observed outcome, that is,
\begin{align*}
\CSTE(\boldsymbol x)  
= \logit\{E(Y \mid \boldsymbol X = \boldsymbol x, Z=1)\}
- \logit\{E(Y \mid \boldsymbol X = \boldsymbol x, Z=0)\}.
\end{align*}
For binary outcomes, the CSTE curve represents the difference in the logarithm of odds ratio (lnOR) between two treatments, and the lnOR is widely used in studies with binary outcomes \cite{MA2020}. 
If we suppose $Y=1$ to be effective and $Y=0$ non-effective, for subject with $\boldsymbol X=\boldsymbol x$, 
$\CSTE(\boldsymbol x) > 0$ indicates that the subject would benefit from the new treatment ($Z=1$) and $\CSTE(\boldsymbol x) < 0$ otherwise.

To accommodate large numbers of covariates in estimating $\CSTE(\boldsymbol x)$, Guo et al. \cite{Guo2021} considered the logit outcome regression model, that is,
\begin{align} \label{eq1}  
\logit\{\mu(\boldsymbol X,Z)\} = g_1(\boldsymbol X^\top \boldsymbol\beta_1)Z + g_2(\boldsymbol X^\top \boldsymbol\beta_2), 
\end{align}
where $\mu(\boldsymbol X,Z) = E(Y\mid \boldsymbol X,Z)$, $g_1(.)$ and $g_2(.)$ are unknown single-valued functions, 
and $\boldsymbol\beta_1 = (\beta_{11}, \ldots, \beta_{1p})^\top$ and $\boldsymbol\beta_2= (\beta_{21}, \ldots, \beta_{2p})^\top$ are the $p$-dimensional vectors of coefficients; {$\boldsymbol\beta_1$ captures the CSTE, while $\boldsymbol\beta_2$ represents the natural level under control}.
Given model \eqref{eq1} and $\boldsymbol X = \boldsymbol x$, the CSTE curve of interest is derived by
\begin{align}
\CSTE(\boldsymbol x) = g_1(\boldsymbol x^\top\boldsymbol\beta_1).\label{eq:cste2}
\end{align}
When $X$ is the single biomarker (i.e., $p=1$), the model considered by Han et al. \cite{KaiShan2017} is equivalent to model \eqref{eq1} by imposing  $\beta_{1} = \beta_{2} = 1$.

To estimate the CSTE curve \eqref{eq:cste2}, we need to estimate $\boldsymbol \beta_1,\boldsymbol \beta_2$ and functions $g_1(.), g_2(.)$.
Following the algorithms in Guo et al. \cite{Guo2021},  when there is a single covariate, the parameters $\beta_1$ and $\beta_2$ are estimated by maximizing the log-likelihood function of model \eqref{eq1}.
When the number of covariates is more than one, $\boldsymbol \beta_1$ and $\boldsymbol \beta_2$ can be estimated by maximizing the penalized log-likelihood function.
(See the detailed log-likelihood functions in equations (2) and (3) in Guo et al. \cite{Guo2021}.)
By controlling the tuning parameter in the penalty function, one could control the sparsity of the estimate of $\beta_{1k}$ for variable selection.
On the other hand, functions $g_1(.)$ and $g_2(.)$ are approximated by the B-spline basis functions.
To derive the simultaneous confidence band (SCB) for the CSTE curve \eqref{eq:cste2}, the spline-backfitted kernel (SBK) technique is applied to estimate $g_1(.)$ again.
Guo et al. \cite{Guo2021} presented the $100(1-\alpha)\%$ SCB for the CSTE curve \eqref{eq:cste2} in Remark 1, and we followed their algorithms to estimate the SCB.


\subsection{CSTE curve for survival outcomes}  

When the outcome of each subject is survival time, which is {commonly a right-censored time-to-event outcome}, the definition of the CSTE curve is given under the potential outcome framework.
When there are two treatment groups, we let $T^z$ be the potential failure time {in the corresponding treatment groups $z\in\{0,1\}$}.
The corresponding potential conditional hazard function is then defined by
\begin{align*}
\lambda^z(t\mid x) = \lim_{\Delta t\to 0}\dfrac{P(t \leq T^z < t+\Delta t\mid  T^z \geq t, X=x)}{\Delta t}.
\end{align*}
In studies with survival outcomes, the treatment effect is usually presented by the logarithm of hazard ratio (lnHR). 
Thus, the CSTE curve for two treatments at some time $t$ is defined by 
\begin{align*}
\CSTE(x,t) = \log\{\lambda^1(t\mid  x)\}-\log\{\lambda^0(t\mid  x)\}.
\end{align*}

Zhou and Ma \cite{Zhou2012} {considered study} with a single covariate of biomarker and two treatments; Ma and Zhou \cite{Ma2017} later extended the method for the case of multiple treatments. 
In the implementation, we consider the setting of a single covariate and multiple treatments.
Let $X$ be the biomarker of subjects and $\boldsymbol Z = (Z_{1},\dots, Z_{K})^\top$ the multi-valued treatment vector, 
where $Z_{k} = 1$ if the subject receives treatment $k~(k=1, \dots, K)$ and $Z_{k} = 0$ otherwise. 
Let $\lambda^k(t\mid  x)$ be the conditional hazard function of the $k$-th treatment, and the corresponding CSTE curve is derived by
\begin{align*}
\CSTE_{k}(x,t) = \log \{\lambda^k(t \mid  x)\}- \log\{\lambda^0(t \mid  x)\}.
\end{align*}
We focus on a special case where $\CSTE_{k}(x,t) \equiv  \CSTE_{k}(x)$, which holds under many well-known models such as the Cox model and the varying-coefficient Cox model.  
To estimate the CSTE curve, the following varying-coefficient proportional hazard regression model is considered:
\begin{equation}  \label{eq2}
\lambda(t\mid X, Z)=\lambda_0(t) \exp\left\{\boldsymbol\beta(X)^\top \boldsymbol Z + g(X)\right\},
\end{equation}
where $\lambda_0(t)$ is an unknown baseline hazard ratio, 
and function $\boldsymbol\beta(.) = (\beta_1(.), \dots, \beta_K(.))^\top$. 
Model \eqref{eq2} implies that 
\begin{align} 
\CSTE_{k}(x) = \beta_{k}(x), \label{eq:cste-surv}
\end{align}
which measures the lnHR of treatment $k$ versus the referential treatment (i.e., treatment 0). 
If $\CSTE_{k}(x)>0$, treatment $k$ has a larger hazard and would accelerate failure; and $\CSTE_{k}(x) <0$, otherwise.  

To estimate $\CSTE_{k}(x)$, the unconfoundedness (i.e., $\boldsymbol Z\perp T^k \mid  X$) is assumed for each treatment $k$. 
Let $T$ be the potential failure time, where $T=\sum_{k=1}^K I(Z_k=1)T^k + I(\sum_{k=1}^K Z_k=0)T^0$; let $C$ be the right-censored time and $Y=\min(T,C)$ the observed time;  
let $\Delta$ be the censoring indicator, where $\Delta=1$ if $Y \leq C$ and $\Delta=0$ otherwise.   
In addition, the assumption that $C\perp T\mid X$ is made.
Given any value of the biomarker $x_0$, the Taylor expansions of $\boldsymbol\beta(.)$ and $g(.)$ in \eqref{eq2} are considered, and they are written as  
\begin{align*}
\boldsymbol\beta(x) &\approx \boldsymbol\beta(x_0)+\boldsymbol\beta'(x_0)(x - x_0)\equiv \boldsymbol\delta+\boldsymbol\gamma (x-x_0) \\
g(x) &\approx  g(x_0)+ g'(x_0)(x - x_0)\equiv g(x_0)+d(x-x_0).
\end{align*}
Thus, estimating the CSTE curve \eqref{eq:cste-surv} is converted to estimating $\boldsymbol\delta$ and the first-order derivatives $\boldsymbol\gamma$ and $d$ \cite{Ma2017,Zhou2012}. 
The parameters $\boldsymbol\delta, \boldsymbol\gamma$ and $d$ can be estimated by maximizing the logarithm of the local partial likelihood function.
(The details of the local partial likelihood function are shown in equation (4) in Ma and Zhou \cite{Ma2017}.)

When there are multiple treatments, in general, we may consider the linear combination of $\CSTE_{k}(x)$ \eqref{eq:cste-surv}.
By introducing the contrast vector $\boldsymbol l = (l_1, \dots, l_K)^\top$, the linear combination of $\CSTE_{k}(x)$ is written by $\boldsymbol l^\top\boldsymbol\beta(x) = l_1\beta_1(x) + \dots + l_K\beta_K(x)$.
For example, when $\boldsymbol l = (1, \dots, 0)^\top$, the linear combination allows one to estimate $\CSTE_{1}(x) = \beta_1(x)$.
Ma and Zhou \cite{Ma2017} presented the simultaneous asymptotic property of $\boldsymbol l^\top\boldsymbol\beta(x)$ and proposed the resampling technique to obtain the SCB. 

\clearpage

\section{Description of the R package {CSTE}}

As mentioned in Section 2.5 of the main text, we developed the R package for estimating $\CSTE(\boldsymbol x)$, the main functions and their descriptions in Table S\ref{cste-pkg}.

\begin{table}[!ht]
\centering
\caption{\label{cste-pkg} Overview of main functions in {CSTE} package.}
\fontsize{9}{11}\selectfont
\begin{tabular}{lll}
\toprule
Outcome & Function & Description \\ \hline
Binary  & \cstebin & Estimate $\CSTE(\boldsymbol x)$ defined in equation (3) in the main text. \\
        & \cstebinSCB &  Estimate the simultaneous confidence band (i.e., SCB).  \\ 
\midrule
Time-to-event   &   \cstesurv &   Estimate $\CSTE(\boldsymbol x)$ defined in equation (4) in the main text. \\
        &  \cstesurvSCB   &  Estimate the SCB.    \\ 
\bottomrule
\end{tabular}
\end{table}  

\clearpage
\section{Screenshots of analyses}

\subsection{Example 1}

\begin{figure}[!htp] 
\centering
\includegraphics[width=\textwidth]{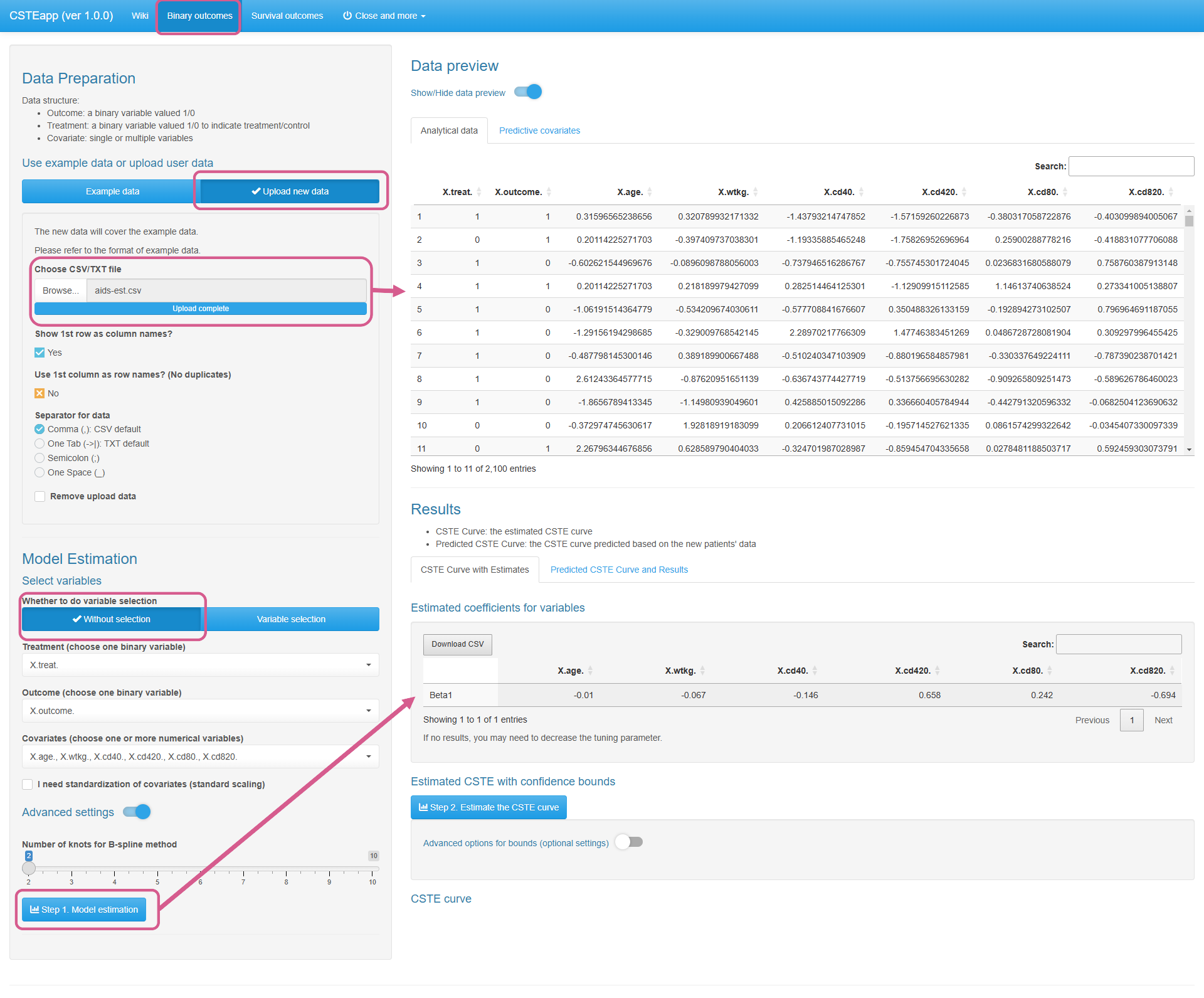}  
\caption{Example 1: Step 1 and 2.}\label{pp1}
\end{figure}

\begin{figure}[!htp] 
\centering
\includegraphics[width=\textwidth]{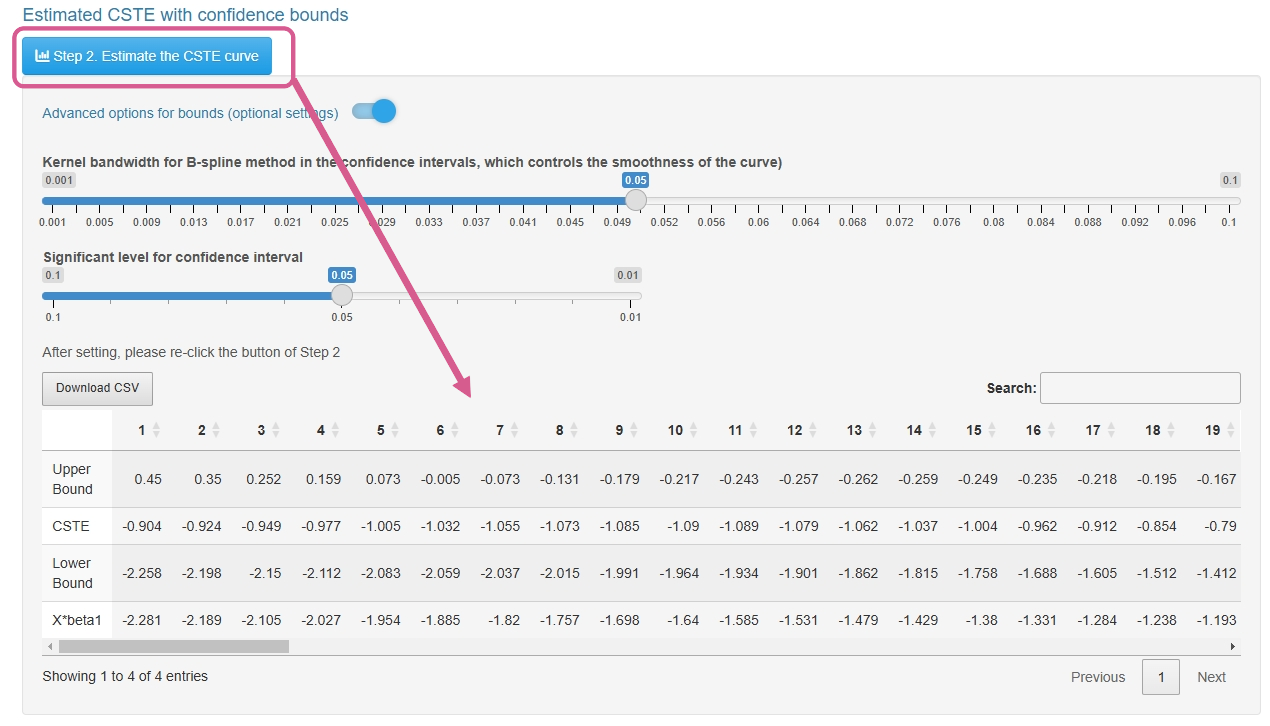}  
\caption{Example 1: Step 3.}\label{pp2}
\end{figure}

\begin{figure}[!htp] 
\centering
\includegraphics[width=\textwidth]{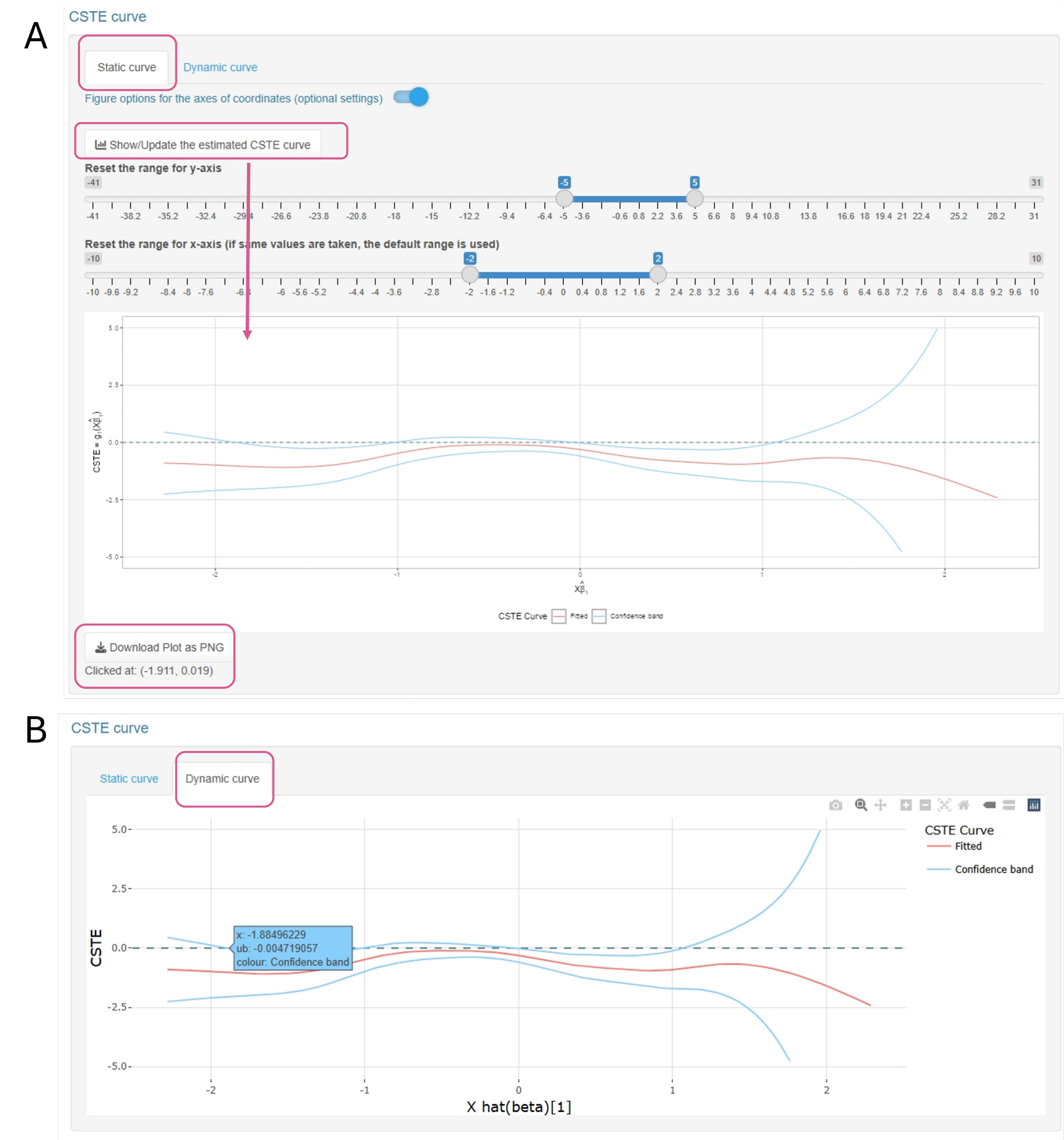}  
\caption{Example 1: Step 4.}\label{pp3}
\end{figure}

\begin{figure}[!htp] 
\centering
\includegraphics[width=\textwidth]{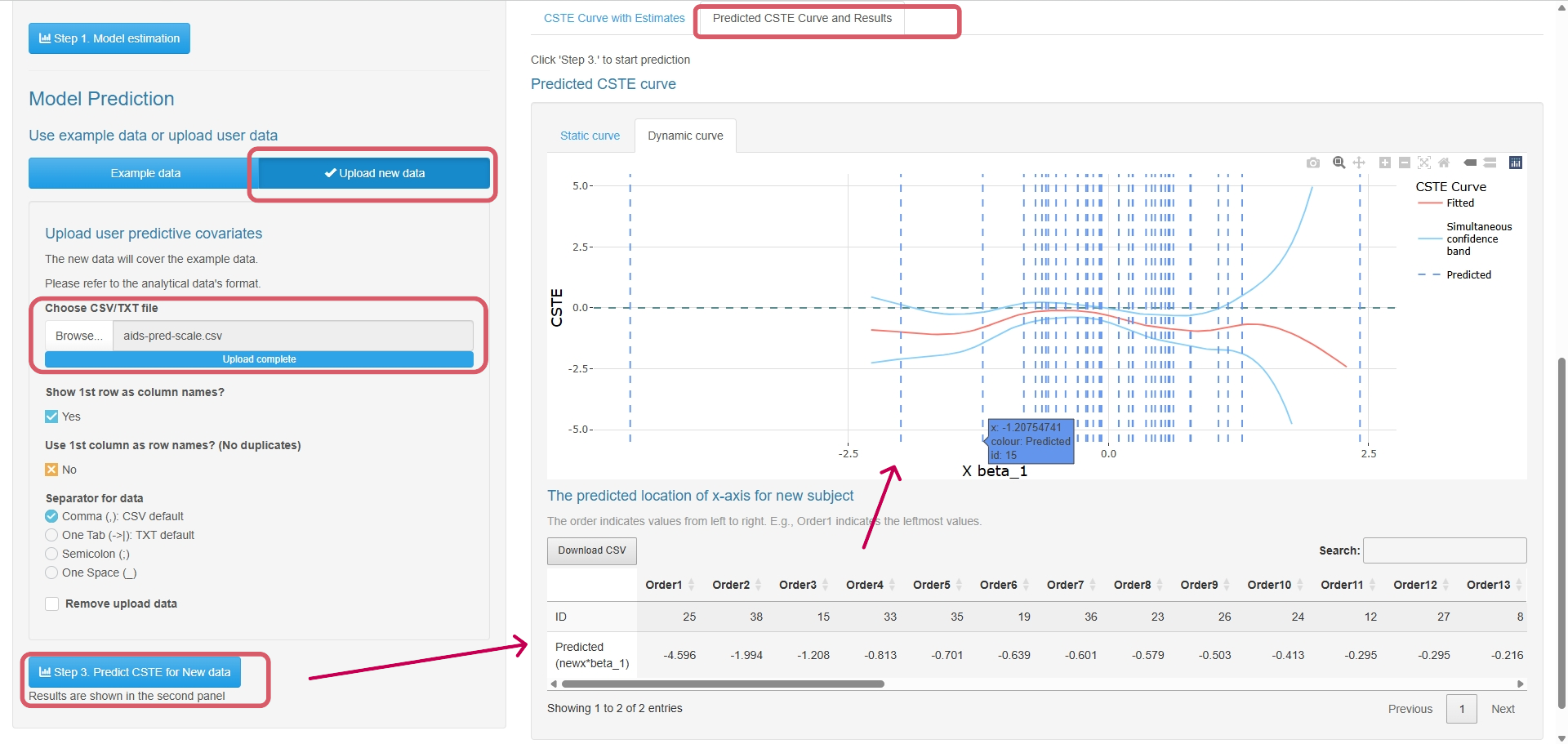}  
\caption{Example 1: Step 5 and 6.}\label{pp4}
\end{figure}

The analytical process is summarized as the following steps (Figures S\ref{pp1}--\ref{pp4}).
\begin{enumerate}[label=Step~\arabic*]
\item Upload data.
\item Set variables and parameters under ``Without variable selection'' and estimate the coefficients.
\item Estimate the CSTE curve.
\item Show the estimated CSTE curve (static plot and dynamic plots); if parameters are changed, plots should be updated. 
\item Upload new data for prediction.
\item Show the predicted values for the new subjects on the CSTE curve (static plot and dynamic plots); if parameters are changed, plots should be updated. 
\end{enumerate}

\clearpage

\subsection{Example 2}

\begin{figure}[!htp] 
\centering
\includegraphics[width=\textwidth]{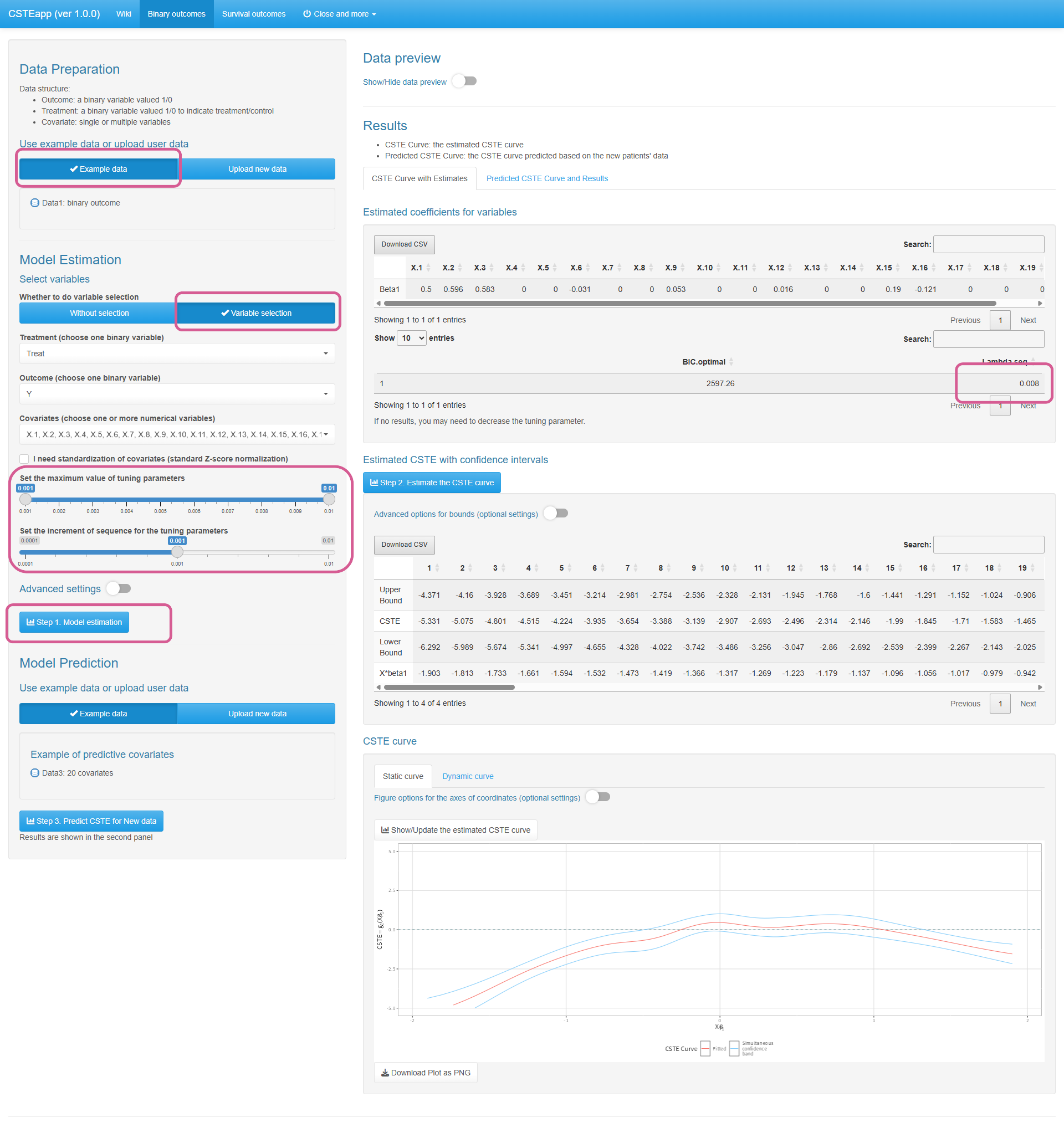}  
\caption{Example 2: Estimation of the CSTE curve (Steps 1--4).}\label{pp5}
\end{figure}

\begin{figure}[!htp] 
\centering
\includegraphics[width=\textwidth]{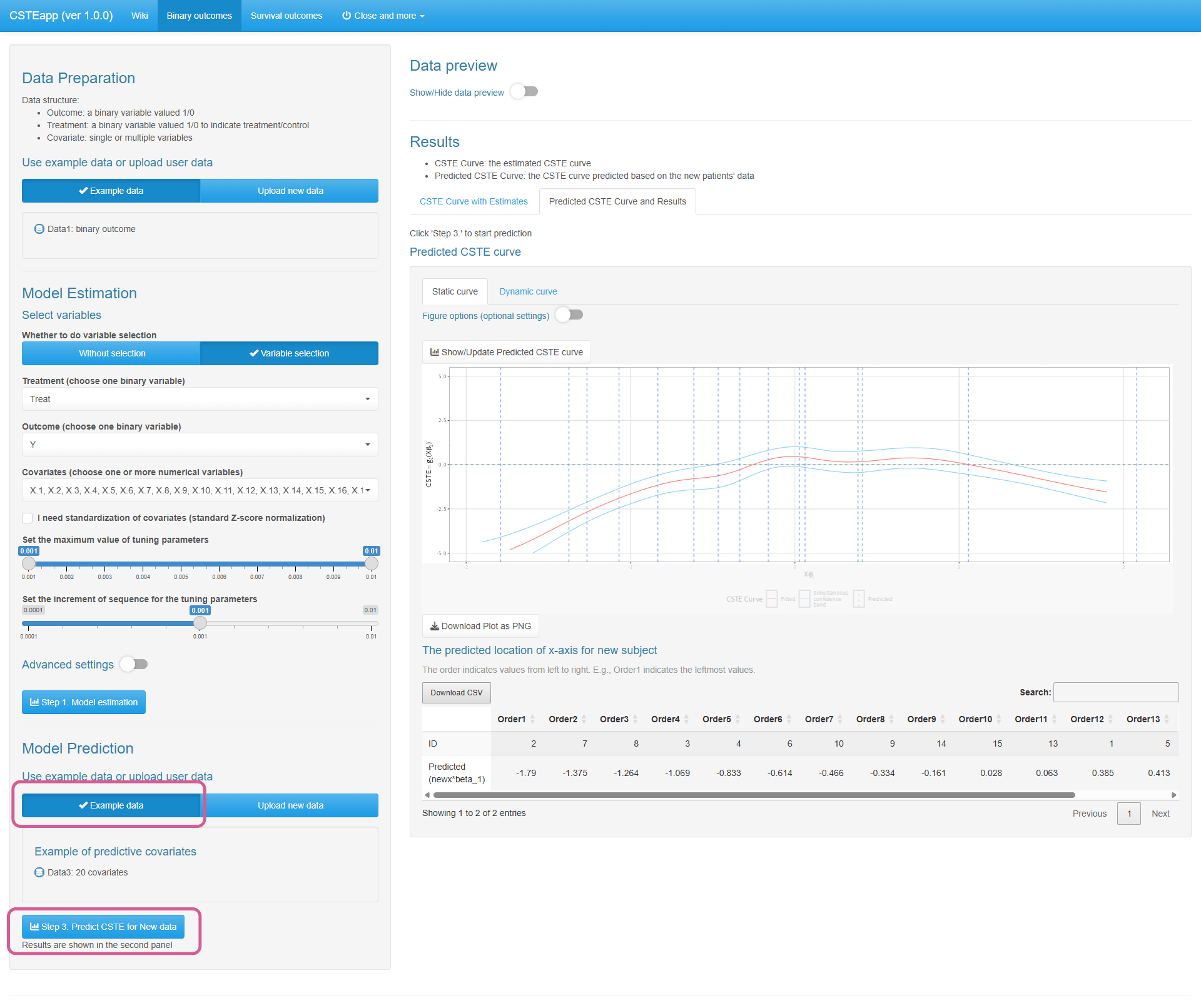}  
\caption{Example 2: Prediction of the CSTE curve (Steps 5--6).}\label{pp6}
\end{figure}

The analytical process is summarized as the following steps (Figures S\ref{pp5}--\ref{pp6}).
\begin{enumerate}[label=Step~\arabic*]
\item Use example data.
\item Set variables and parameters under ``Variable selection'' and estimate the coefficients.
\item Estimate the CSTE curve.
\item Show the estimated CSTE curve (static plot and dynamic plots); if parameters are changed, plots should be updated. 
\item Use example data for prediction.
\item Show the predicted values for the new subjects on the CSTE curve (static plot and dynamic plots); if parameters are changed, plots should be updated. 
\end{enumerate}

\subsubsection{Validation of estimation results}
As mentioned at the end of Section 3.2, we present the details of the data-generating process and the validation of estimations.

For data with binary outcomes, we simulated 2000 samples and 20 covariates $\boldsymbol X$, where $\boldsymbol X$ was simulated from the truncated multivariate normal distribution with mean 0, covariance matrix $\boldsymbol\Sigma_{ij} = 0.5^{|i-j|}$, and truncated by $(-2,2)$. 
The binary treatment variable $Z$ was sampled from the binomial distribution $Bin(1,0.5)$.
To generate the outcomes, the following true model is considered: that is,
\begin{align*}
\logit(\mu(\boldsymbol X,Z)) = \boldsymbol X^\top \boldsymbol\beta_1(1 - \boldsymbol X^\top \boldsymbol\beta_1)Z + \exp(\boldsymbol X^\top \boldsymbol\beta_2),
\end{align*}
where $\boldsymbol\beta_1$ and $\boldsymbol\beta_2$ were set to be $(1,1,1,0,\dots,0)^\top/\sqrt{3}$ and $(1,-2,0,\dots,0)^\top/\sqrt{5}$, respectively.  
Then, each binary outcome $Y$ was simulated from $Bin(1,\mu(X,Z))$.

\begin{figure}[!ht]
\centering
\includegraphics[width=\textwidth]{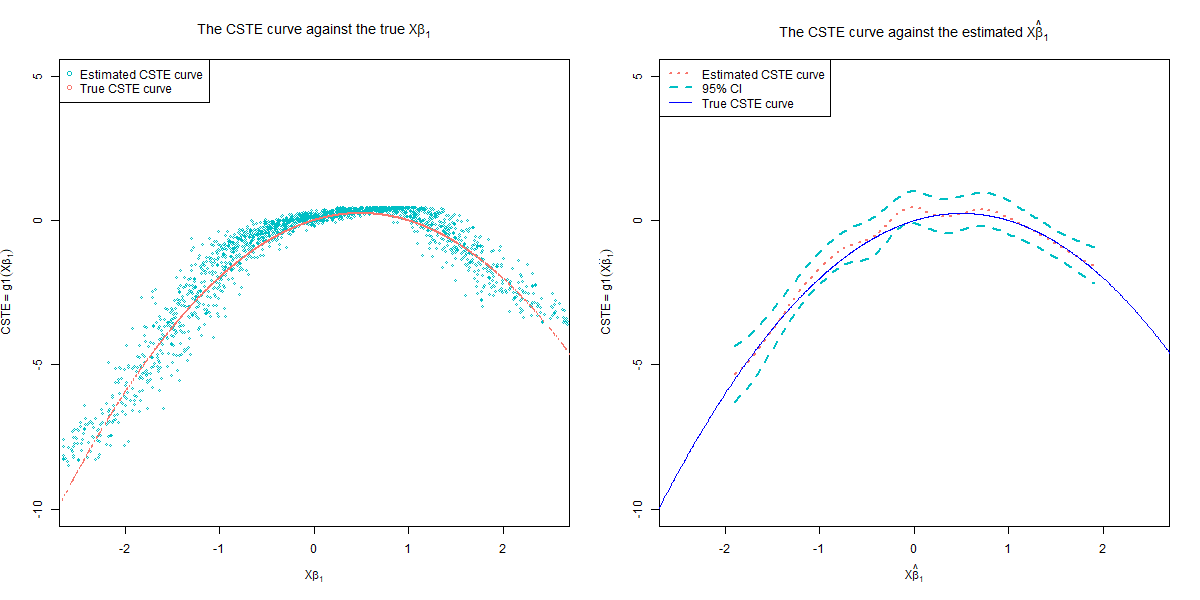}  
\caption{\label{fig1} Example 2: B-spline point estimates of CSTE curve.}
\end{figure}

The true CSTE points and the corresponding B-spline point estimates were presented as the red and green points, respectively, in the left panel of Figure S\ref{fig1}. 
The spline-backfitted kernel (SBK) estimate, $\hat g_{1,skb}(\boldsymbol x\hat{\boldsymbol \beta}_1)$, of the CSTE curve and its SCB was shown in the right panel of Figure S\ref{fig1}. 
The estimated CSTE curve was close to the true value.

\clearpage
\subsection{Example 3}

\begin{figure}[!htp] 
\centering
\includegraphics[width=\textwidth]{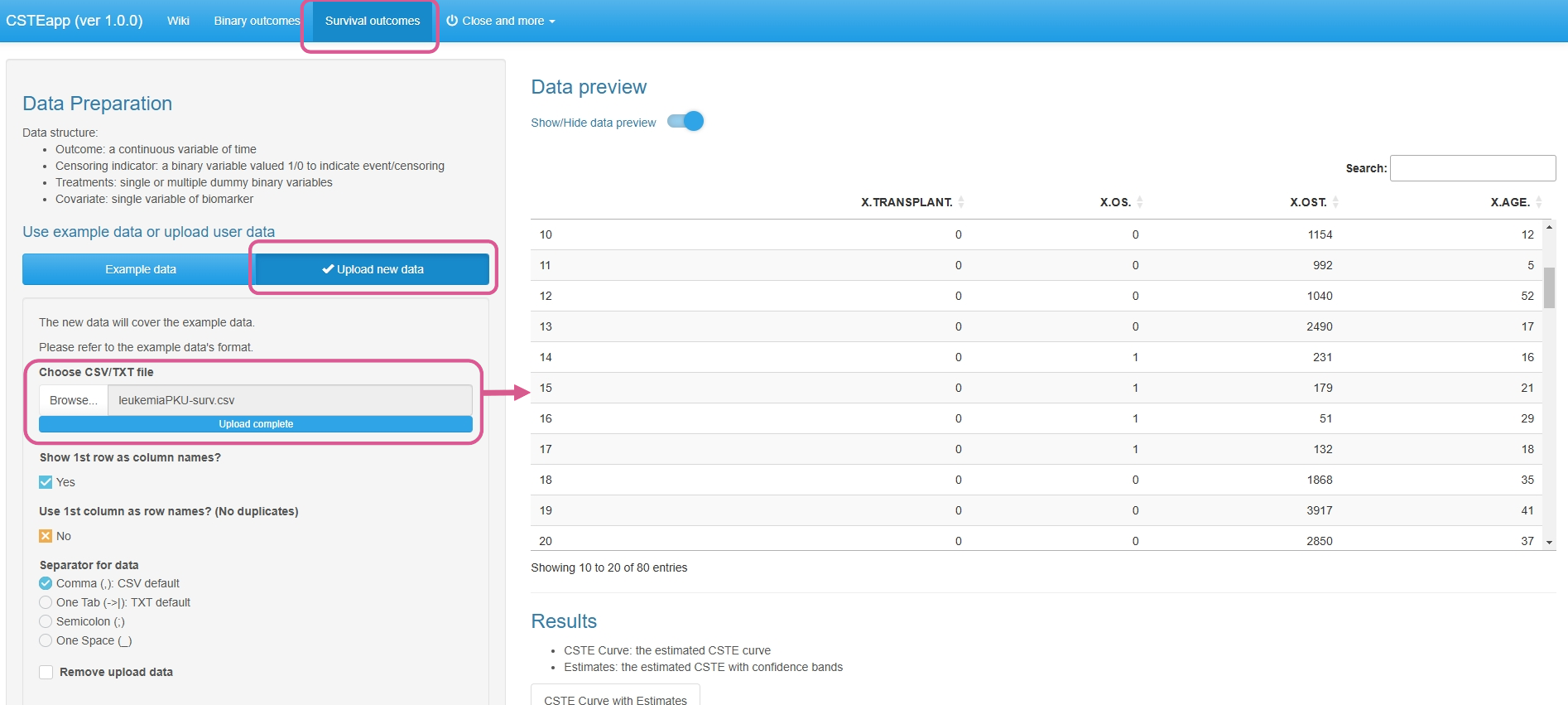}  
\caption{Example 3: Upload data (Step 1).}\label{pp7}
\end{figure}

\begin{figure}[!htp] 
\centering
\includegraphics[width=\textwidth]{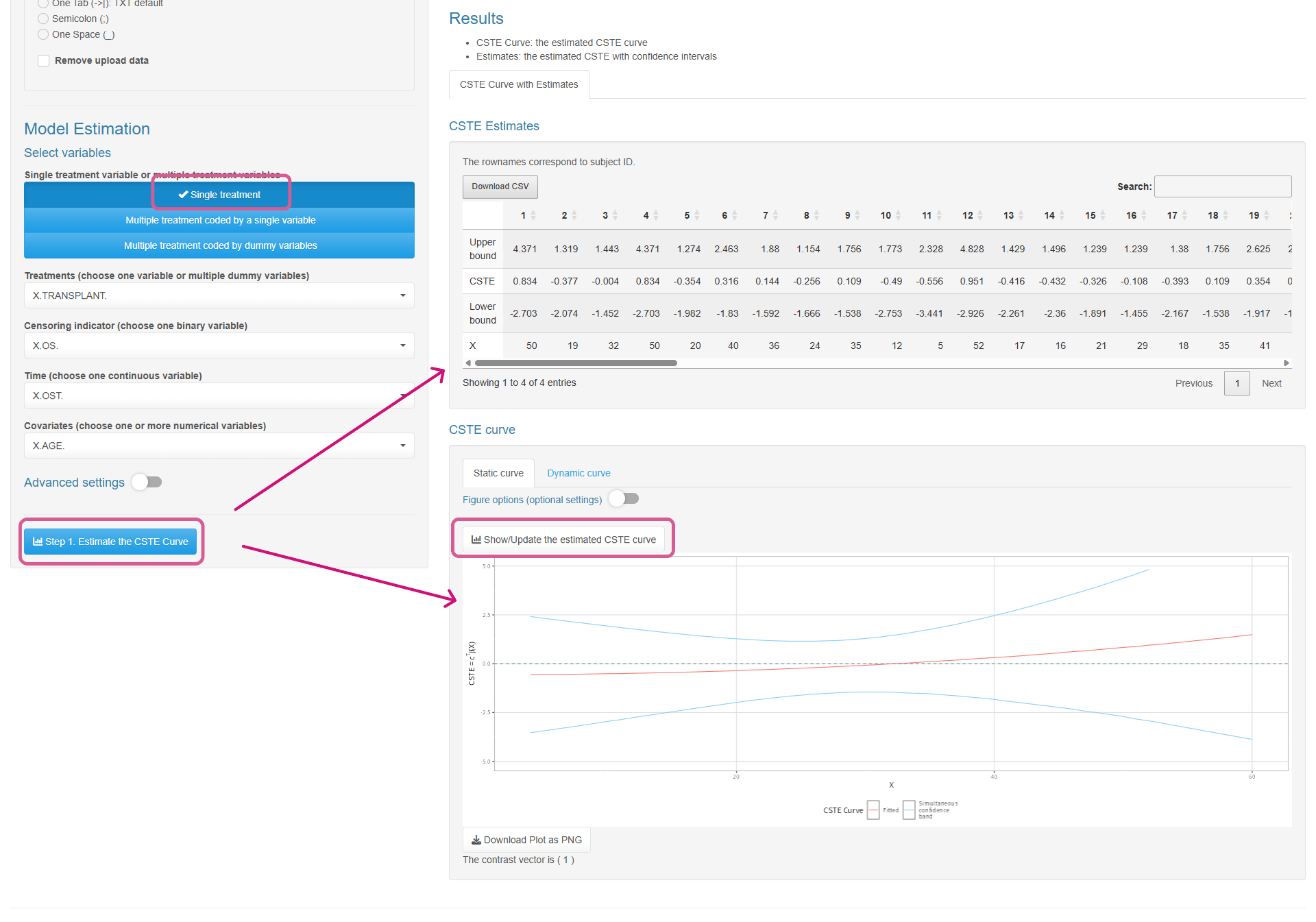}  
\caption{Example 3: Set variables and parameters and estimate the CSTE curve (Steps 2--3).}\label{pp8}
\end{figure}

The analytical process is summarized as the following steps (Figures S\ref{pp7}--\ref{pp8}).
\begin{enumerate}[label=Step~\arabic*]
\item Upload local data.
\item Set variables and parameters under ``Single treatment'' and estimate the CSTE curve.
\item Show the estimated CSTE curve (static plot and dynamic plots); if parameters are changed, plots should be updated. 
\end{enumerate}

\clearpage
\subsection{Example 4}

\begin{figure}[!htp] 
\centering
\includegraphics[width=\textwidth]{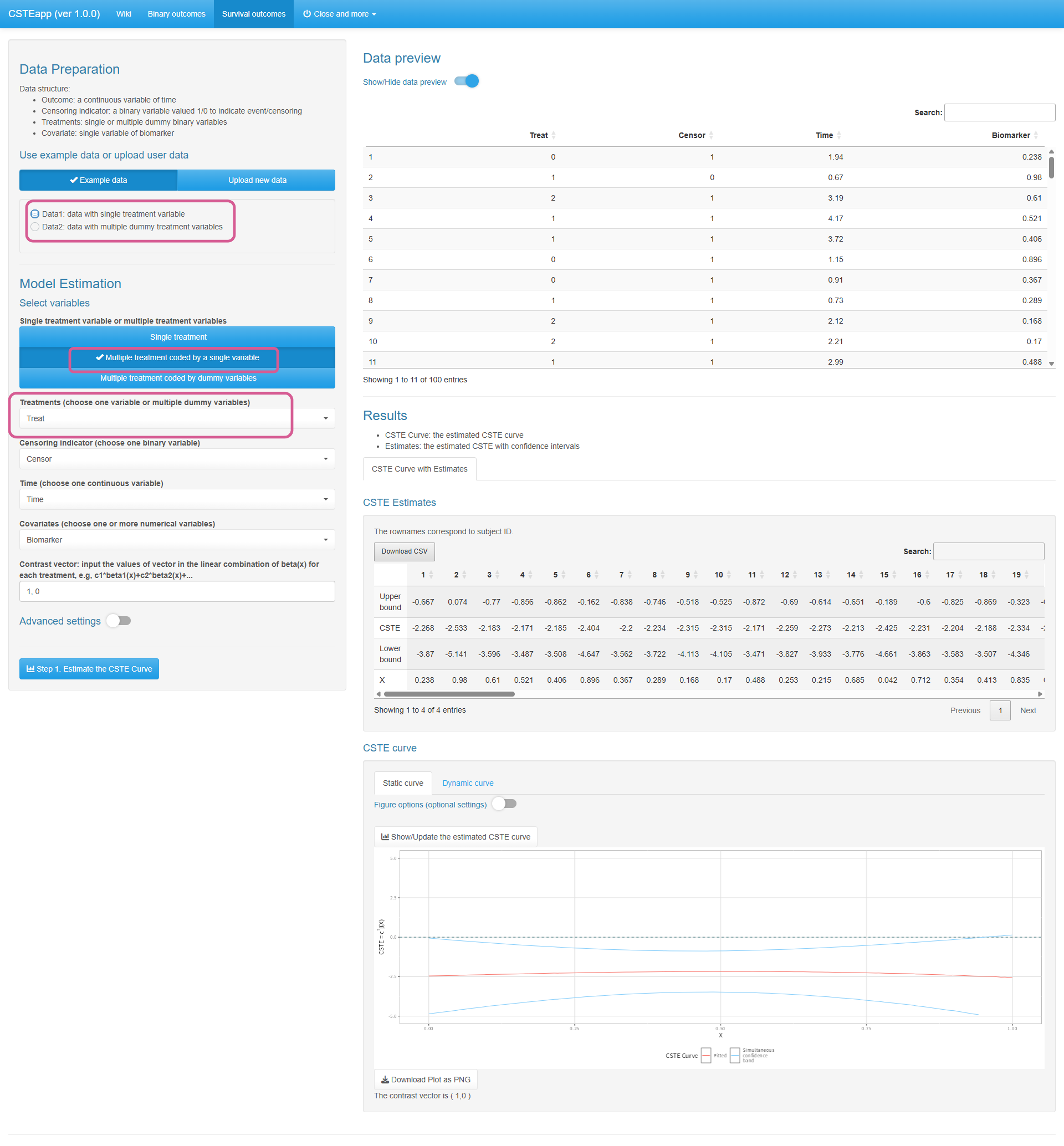}  
\caption{Example 4: Estimation of the CSTE curve based on Data 1.}\label{pp9}
\end{figure}

\begin{figure}[!htp] 
\centering
\includegraphics[width=\textwidth]{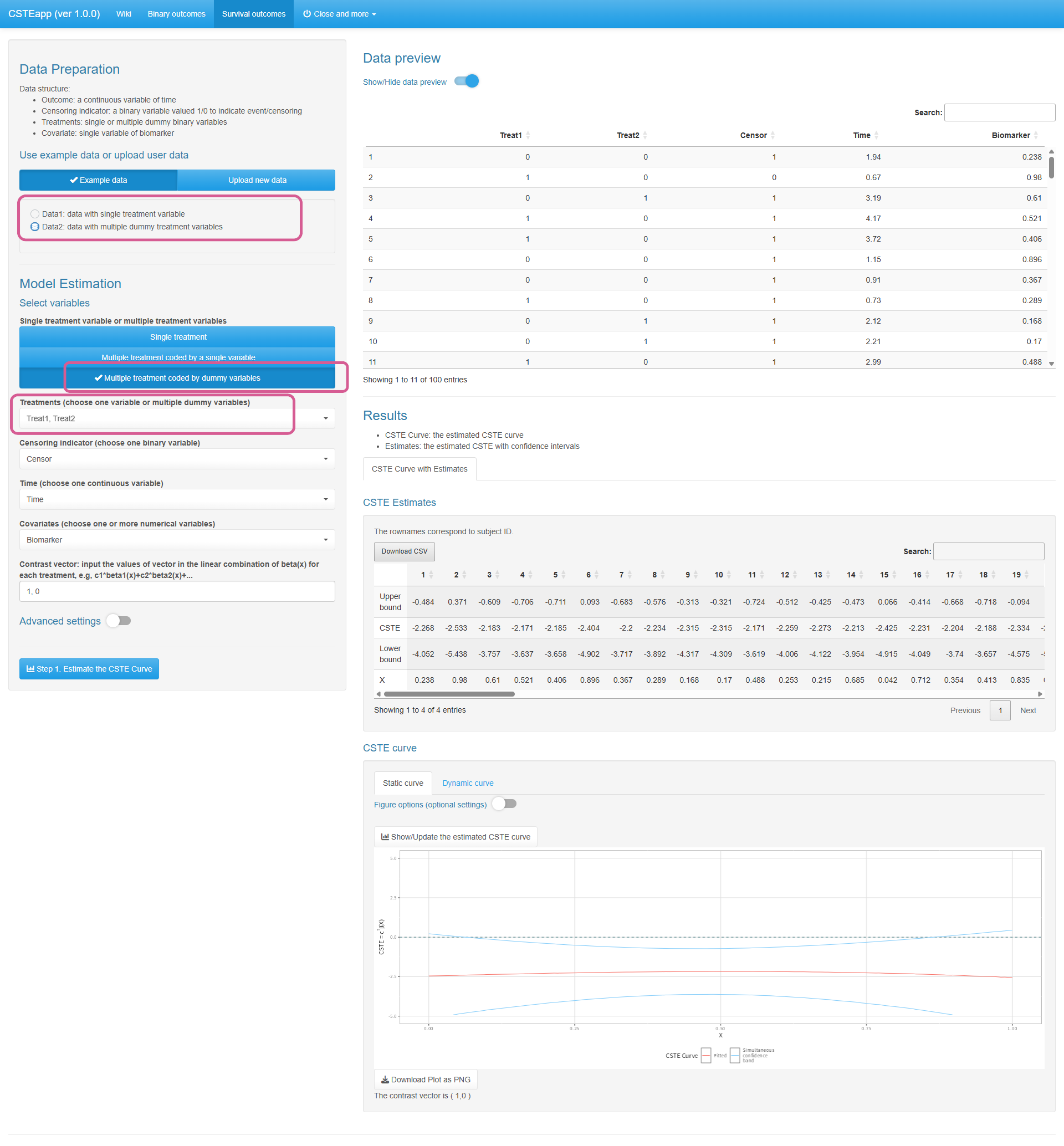}  
\caption{Example 4: Estimation of the CSTE curve based on Data 2.}\label{pp10}
\end{figure}

The analytical process is summarized as the following steps (Figures s\ref{pp9}--\ref{pp10}).
\begin{enumerate}[label=Step~\arabic*]
\item Choose example data.
\item If Data 1 is chosen, then set variables and parameters under ``Multiple treatment coded by a single variable''; if Data 2 is chosen, then set variables and parameters under ``Multiple treatment variable''; input the contrast vector, and estimate the CSTE curve.
\item Show the estimated CSTE curve (static plot and dynamic plots); if parameters are changed, plots should be updated. 
\end{enumerate}

\clearpage
\subsubsection{Validation of estimation results}
As mentioned in Section 3.4, we present the data-generating process.

In {Example 4}, we considered 100 samples. 
The single random variable of biomarker $X$ was sampled from the uniform distribution $U[0,1]$. 
The multiple treatment $Z$ consisted of two binary variables $(Z_{1}, Z_{2})$, 
where $Z_{1}$ followed the Bernoulli distribution $Ber(0.3)$ and $Z_{2} = b\times (1 - Z_{1})$ with $b \sim Ber(0.5)$.
Let $\boldsymbol\beta(x) = \left(\beta_{1}(x), \beta_{2}(x)\right)$, where $\beta_{1}(x) = -1 - \exp(x)$ and $\beta_{2}(x) = -\exp(x)$.  
The true model is set up with the following hazard function:
\begin{align*}
\lambda(t|  X, \boldsymbol Z)=0.6t^2 \exp\left\{\boldsymbol\beta(X)^\top \boldsymbol Z + X^2\right\}. 
\end{align*}
The censoring time was sampled from the exponential distribution with intensity $\lambda=0.23X$, which was adjusted such that the censoring rate is approximately $20\%$.

\begin{figure}[!ht] 
\centering
\includegraphics[width=\textwidth]{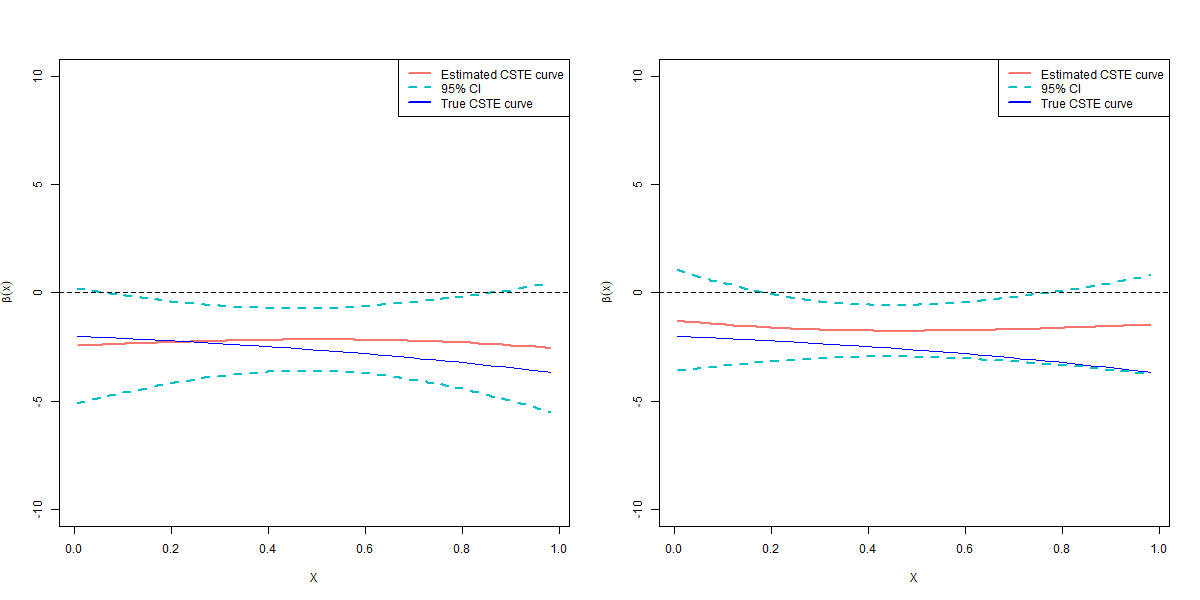} 
\caption{\label{fig2} Example 4: comparison of the estimated CSTE curve and the true CSTE curve.}
\end{figure}

To validate the results, we compare the estimation with the true setting in data generation.
Using the same settings of parameters in the main text, we considered estimating the CSTE curve with $\boldsymbol l^\top = (1, 0), (0,1)$, which were presented in the left and right panels of Figure S\ref{fig2}, respectively.

The reproducible R codes are presented in \url{https://github.com/mephas/CSTEapp-Rcodes}.

\end{document}